\documentclass[preprint]{aastex63}
\usepackage{amsmath,amstext}
\usepackage[T1]{fontenc}
\usepackage{apjfonts} 
\usepackage[figure,figure*]{hypcap}

\received{~--- xxx}
\revised{~--- xxx}
\accepted{~--- xxx}

\shorttitle{Compact star-forming galaxies: Blueberry}
\shortauthors{Paswan et al.}

\begin{document}

\title{\large Unveiling an old disk around a massive young leaking blueberry in SDSS-IV MaNGA}

\correspondingauthor{Abhishek Paswan}
\email{paswanabhishek@iucaa.in}

\author{Abhishek Paswan}
\affil{Inter-University Centre for Astronomy and Astrophysics, Ganeshkhind, Post Bag 4, Pune 411007, India}

\author{Kanak Saha}
\affiliation{Inter-University Centre for Astronomy and Astrophysics, Ganeshkhind, Post Bag 4, Pune 411007, India}

\author{Anshuman Borgohain}
\affiliation{Department of Physics, Tezpur University, Napaam 784028, India}

\author{Claus Leitherer}
\affiliation{Space Telescope Science Institute, Baltimore, MD 21218, USA}

\author{Suraj Dhiwar}
\affiliation{Inter-University Centre for Astronomy and Astrophysics, Ganeshkhind, Post Bag 4, Pune 411007, India}
\affiliation{Department of physics, Savitribai Phule Pune University, Pune 411007, India}

\begin{abstract}

Extreme emission-line galaxies, such as blue compact dwarfs (BCDs), Green Peas (GPs) and blueberries in the local Universe are the potential candidates to understand the nature of galaxies that re-ionized the early Universe. Being low-mass, metal-poor starburst systems they are understood as local analogs of the high redshift Lyman Continuum (LyC) and Lyman-$\alpha$ emitters (LAEs). Even with their proximity to us, we know little about their spatially resolved properties, most of the blueberries and GPs are indeed compact, remain unresolved. Here, we report the detection of a disk-like lower-surface brightness (LSB) stellar host with very old population around a blueberry LAE system using broad $i$-band imaging and integral field spectroscopic data from SDSS and SDSS-IV MaNGA surveys, respectively. The LSB stellar host is structurally similar to that observed around local starburst BCDs. Furthermore, the kinematics of the studied blueberry source bear sign of misalignment between the gas and stellar components. Our findings establish an intriguing thread connecting the blueberry and an LSB disk with old stellar population, and suggest that blueberries and their high redshift counterparts such as GPs do not represent peculiar cases of dwarf galaxy evolution. In fact, with respect to the structural properties of their host galaxies, they are  compatible with a common evolutionary track of the main population of local BCDs.

\end{abstract}

\keywords{Galaxy evolution --- Starburst galaxies --- H{\sc ii} regions --- Interstellar medium  --- Radiative transfer --- Reionization}

\section{Introduction} \label{sec:intro}

Compact, emission-line, low-mass star-forming (CELLs) galaxies with high ionization parameter (characterized by large values of [O{\sc iii}]/[O{\sc ii}] $\gtrsim$ 5) and low oxygen abundance (7.1 $\leqslant$ 12 + log(O/H) $\leqslant$ 8.6 $\equiv$ metallicity; hereafter) such as Green Peas \citep[GPs;][]{Cardamoneetal2009} and blueberries \citep[the nearby counterparts of GPs;][]{Yangetal2017} are now believed as one of the potential sources of escaping Lyman Continuum \citep[LyC;][]{Izotovetal2016, Izotovetal2018} and Lyman-$\alpha$ \citep[Ly$\alpha$;][]{Jaskot2019} photons in the local Universe. Their high redshift analogs are thought to have played a key role in the re-ionization of our early Universe \citep{Shapleyetal2016, Bianetal2017, Vanzellaetal2018, Fletcheretal2019, Sahaetal2020}. Although there has been a tremendous progress enhancing our knowledge of these galaxies as LyC and Ly$\alpha$ leakers \citep{NakajimaOuchi2014, Verhamme2017, Kimmetal2017, Trebitschetal2017, Trebitschetal2020, Izotovetal2018b, Wangetal2019,Huan2017}, little is known about their whereabouts - especially the morphological structure of stellar and gas components, composition of the underlying stellar populations, star formation history (SFH), kinematics of stars and gas and finally their origin. \\
\par
The extreme emission line properties, as noticed in the case of GP and blueberry systems, have also been observed in extremely metal-poor blue compact dwarf (XBCD) galaxies \citep{Terlevich1991,Izotov1997,Papaderos1998,Papaderos2008}, including a few ultra-compact dwarf galaxies \citep[e.g.,][]{Reverte2007}. The locations of GPs on the fundamental plane (i.e., the mass-metallicity and luminosity-metallicity relations) are found systematically offset to lower metallicities, compared to the majority of local star forming galaxies \citep[SFGs;][]{Amorin2010,Amorin2011}. This implies that GPs and their local counterparts (i.e., blueberries) most likely form a distinct class of objects, along with local XBCDs \citep{Guseva2009} and a few luminous BCDs at low and intermediate redshifts \citep{Ostlin2001,Hoyos2005,Kakazu2007,Salzer2009} and most SFGs at high redshift \citep{Pettini2001,Perez-montero2009,Finkelstein2011}. Based on their extreme emission line properties, \citet{Cardamoneetal2009} and \citet{Yangetal2017} suggested that GPs and blueberries could potentially be classified as part of the heterogeneous luminous BCD category. Furthermore, the work by \citet{Izotov2011} also showed that GPs are a subset of luminous compact galaxies whose metallicties are similar to low luminosity BCDs. However, it remains unclear whether the GPs and blueberries have their morphological structure and SFHs similar to luminous BCDs. It is well known that XBCDs and some luminous BCDs possess a compact to moderately extended stellar host underlying their recent star formation component \citep[][ and references therein]{Papaderos2008}. This indicates that these systems have not experienced in-situ formation of their first generation of stellar population in a galaxy-wide starburst. In fact, their exponential light profile at the outskirt and the structural properties of their host are found fairly comparable to the bona fide old and metal-rich BCDs, containing a low-mass fraction of very old ($\sim$ 10 Gyr) stars as well \citep[e.g.,][]{Kunth1988,Papaderos1996a,Papaderos1996b,Cairos2001,Noeske2003,GildePaz2005,Guseva2004,Papaderos2008}. Such morphologies, stellar population and SFHs in the case of GPs and blueberries remain less explored. The existence of an old stellar population, in particular in the low-mass end of the galaxy mass spectrum (i.e., the CELLs galaxies), may shed further light on our current understanding of the formation of disks and stellar feedback \citep{MoMaoWhite1998,FerraTolstoy2000,Salesetal2009,Dutton2009,Agertzetal2011,Ubleretal2014,RathausSternberg2016}.\\

\par
In the recent past, most CELLs galaxies (i.e., GPs and blueberries) are being discovered and studied using data from the Sloan Digital Sky Survey \citep[SDSS;][]{Lintott2008,Lintott2011}, where this survey provides us with the seeing-limited ($\sim$ 1.2" FWHM) images and fiber-slit \citep[3'' in diameter;][]{Yorketal2000} based spectra with poor signal-to-noise-ratio (S/N) and shallow sensitivity \citep{Abazajian2009,Blanton2017}. As a result, the CELLs galaxies remain unresolved in the SDSS images, preventing one to further explore their spatially-resolved morphological structure. The similar status remain prevailed on the spectral side. So far, it has not become possible to find a direct observational evidence of old stellar population in these galaxies. At the same time, SDSS spectra does not allow one to carry out the spatially resolved kinematics of stars and gas. A number of efforts have been made to observe these galaxies using the Hubble Space Telescope (HST) Cosmic Origins Spectrograph (COS) in the ultra-violet (UV) domain \citep[e.g.,][]{Izotovetal2016,Izotovetal2018,Izotovetal2018b,Rong2018,Malkan2019,kim2021,Izotov2021}. These observations have brought us significant information on the UV morphologies that are characterized by the bright star-forming knots in the central region of some of these GPs and blueberries. Although these observations (mostly tracing the spatial distribution of young stars) reveal the presence of possible exponential UV disks having scale lengths in the range of $0.6 - 1.4$ kpc, the results are, however, hampered by the limited unvignetted aperture of the HST/COS spectrograph. \\

\par
In order to overcome the issues with spatial resolution and spectroscopic sensitivity for studying the morphological structure, stellar population and SFHs of the GP galaxies, \citet{Amorinetal2012} for the first time presented three GPs observed with $HST$ $R$-band imaging and long-slit deep optical spectroscopy using OSIRIS instrument mounted on the 10.4-m GTC. In their spatially-resolved images of GPs, they found that these systems seem to have compact ($\sim$ 5 kpc) and irregular morphology of their high surface brightness components, similar to typical BCDs. Furthermore, they also reported the presence of a Lower Surface Brightness (LSB) envelope with an exponentially decreasing intensity at the outskirt, presumably due to an underlying old stellar population. But they could not rule out if this was a generic property of extended nebular halos that are expected to be present in dwarf galaxies with strong starburst events. In one of the GPs, they did provide a direct evidence for the old stellar population, through the detection of a weak MgI $\lambda$5173 absorption line \citep[GP113303;][]{Amorinetal2012}. Overall, their study concludes that GPs are old galaxies with most of the stellar masses formed several Gyrs ago. A similar conclusion has also been reached in a recent work by \citet{Clarkeetal2021} with HST imaging of nine GPs in F555W and F850LP filters. Although stellar kinematics of GPs remain largely unexplored, gas kinematics was presented in four GPs using  Integral Field Unit (IFU) spectroscopic observations \citep{Lofthouse2017}. Overall, the underlying physical nature of the CELLs galaxies is still limited and no doubt of a need for further exploration with promising IFU observations.\\  

\par
In the current work, we present a CELLs galaxy (viz. SHOC~579 or MaNGA ID: $8626 - 12704$, RA: 17$^{h}$ 35$^{m}$ 01.25$^{s}$ and DEC: +57$^{d}$ 03$^{m}$ 09$^{s}$, see left panel in Fig.~\ref{fig:1}) observed with optical IFU spectroscopy at the Mapping Nearby Galaxies at Apache Point Observatory \citep[MaNGA;][]{Bundyetal2015}. SHOC~579 is the only nearby Ly$\alpha$ emitting galaxy (z $\sim$ 0.0472) which is observed in the MaNGA IFU survey. The escape fraction of Ly$\alpha$ photons is measured to be $f^{ecp}_{Ly\alpha}$ $\sim$10\% \citep[see,][]{Jaskot2019}. The close proximity of SHOC~579 in combination with its observed MaNGA IFU data having a better S/N compared to single fiber-slit based SDSS data provide us an unique opportunity to address the above mentioned missing properties of blueberry galaxy in unprecedented detail.\\ 

\par
Throughout the paper, we have considered a flat $\Lambda$CDM cosmology with $H_{0}$ = 70 km$s^{-1}$ Mpc$^{-1}$ , $\Omega_{m}$ = 0.3, and $\Omega_{\Lambda}$ = 0.7, where $H_{0}$ represents the Hubble constant, and $\Omega_{m}$ and $\Omega_{\Lambda}$ are matter and dark energy density, respectively. All magnitudes quoted in the paper are in AB system \citep{Oke1974}.

\section{Data and galaxy selection}
\label{data}

The data used in the present study come from the 16$^{th}$ data release (DR16) of MaNGA survey. This survey is an optical Integral Field Unit (IFU) spectroscopy observing program under fourth generation of SDSS survey \citep[SDSS-IV;][]{Bundyetal2015}. It uses the BOSS spectrograph \citep{Smeeetal2013} mounted on 2.5-m Sloan Foundation Telescope \citep{Gunnetal2006} at Apache Point Observatory (APO). Here, the selected IFU sizes are such that it covers at least 1.5 R$_{e}$ of the observed galaxies \citep{Lawetal2016}. This survey is targeted to observe $\sim$10,000 nearby (0.01 $\textless$ z $\textless$ 0.15) galaxies having stellar mass $\geqslant$ 10$^{9}$ M$_{\odot}$ \citep{Wake2017}. In this survey, the spectra cover a wavelength range of $3600 - 10300$ \AA, with a spectral resolution of $R$ $\sim$ 2000 and velocity resolution of $\sigma$ $\sim$ 60 km~s$^{-1}$. The observed raw IFU datacube is first reduced and calibrated using Data Reduction Pipeline \citep[DRP;][]{Lawetal2016}, and then analysed after running Data Analysis Pipeline \citep[DAP;][]{Westfalletal2019} over DRP products. DAP uses pPXF code \citep{Cappellari2004} with MILES stellar libraries to fit both the spectral and continuum spectrum simultaneously, and provides spectral line fluxes and their EW maps including gas and stellar kinematics (e.g., $v_{stellar}$, $\sigma_{stellar}$, $v_{gas}$ and $\sigma_{gas}$) etc. Note that all the derived 2D maps of various physical parameters have an effective spatial resolution of 2.5" full width at half maximum (FWHM). All the emission line fluxes used in this study are corrected for both the Galactic and internal extinctions. First, Galactic reddening is applied by assuming reddening law provided by \citet{ODonnell1994}. Then internal reddening correction to the galaxy is applied using flux ratio of f$_{H\alpha}$/f$_{H\beta}$ by assuming its theoretical value as 2.86 and Case-B recombination \citep{OsterbrockBochkarev1989} with an electron temperature of $\sim$ 10$^{4}$ K and electron density of 100 cm$^{-3}$. For some spaxels, the flux ratio of f$_{H\alpha}$/f$_{H\beta}$ is found less than the theoretical value of 2.86. A low value of f$_{H\alpha}$/f$_{H\beta}$ is often associated with intrinsically low reddening \citep{Paswan2018,Paswan2019}, and hence we assumed an internal $E(B - V)$ values as zero for such cases.\\

Apart from the MaNGA data, other ancillary data are used from several publicly available ground and space-based sky surveys such as $GALEX$, SDSS, 2MASS and $Spitzer$. These data are further analyzed using standard packages available in Python \citep{Van2009}, SExtractor \citep{Bertin1996}, GALFIT \citep{Pengetal2002} and IRAF \citep{Tody1986}, wherever required. \\

The galaxy under this study is selected from our search program of blueberry candidates in DR16 of MaNGA survey. In order to select the potential candidates, we applied the following criteria to each galaxy in the entire MaNGA sample: 

\noindent (i) O$_{32}$ $\gtrsim$ 5\\  
\noindent (ii) EW(H$\beta$) $\gtrsim$ 100 \AA, \\ 
\noindent (iii) EW([OIII] $\lambda$5007) $\gtrsim$ 500 \AA.\\ 
\par
By applying criterion $(i)$, we found a total of 50 galaxies. Out of these 50, we found only three potential blueberry candidates that satisfied the last two criteria $(ii, iii)$. Of these three candidates, two are already presented in \citet{Paswan2021}. These two blueberries have been found in close association with Low Surface Brightness disk galaxies, and they are found to be in an advanced state of merger. The remaining galaxy, known as SHOC 579, is studied here in detail.\\

\section{Basic properties of the selected blueberry galaxy}
\label{char}

\begin{figure*}
\begin{center}
\rotatebox{0}{\includegraphics[width=0.44\textwidth]{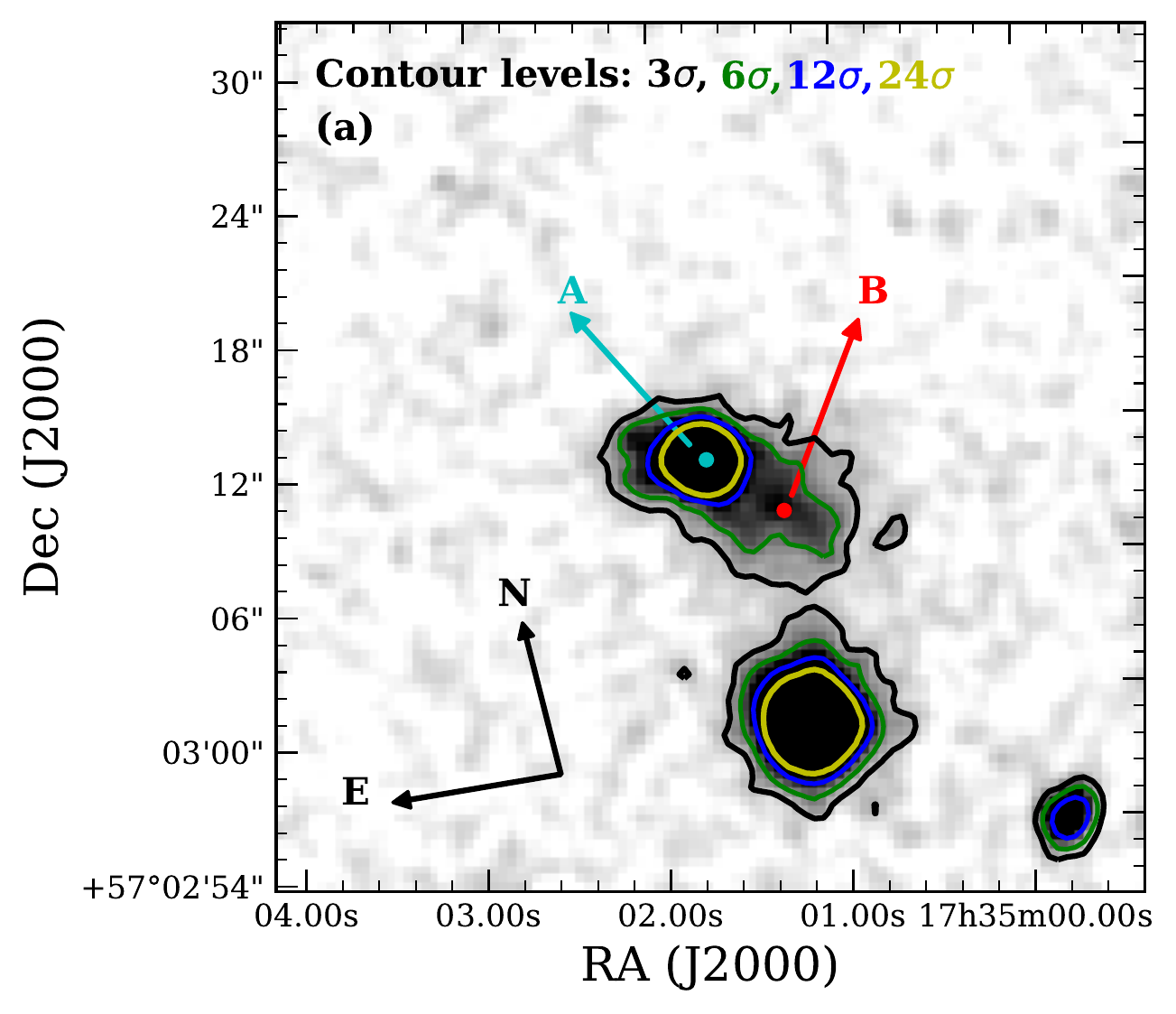}}
\rotatebox{0}{\includegraphics[width=0.45\textwidth]{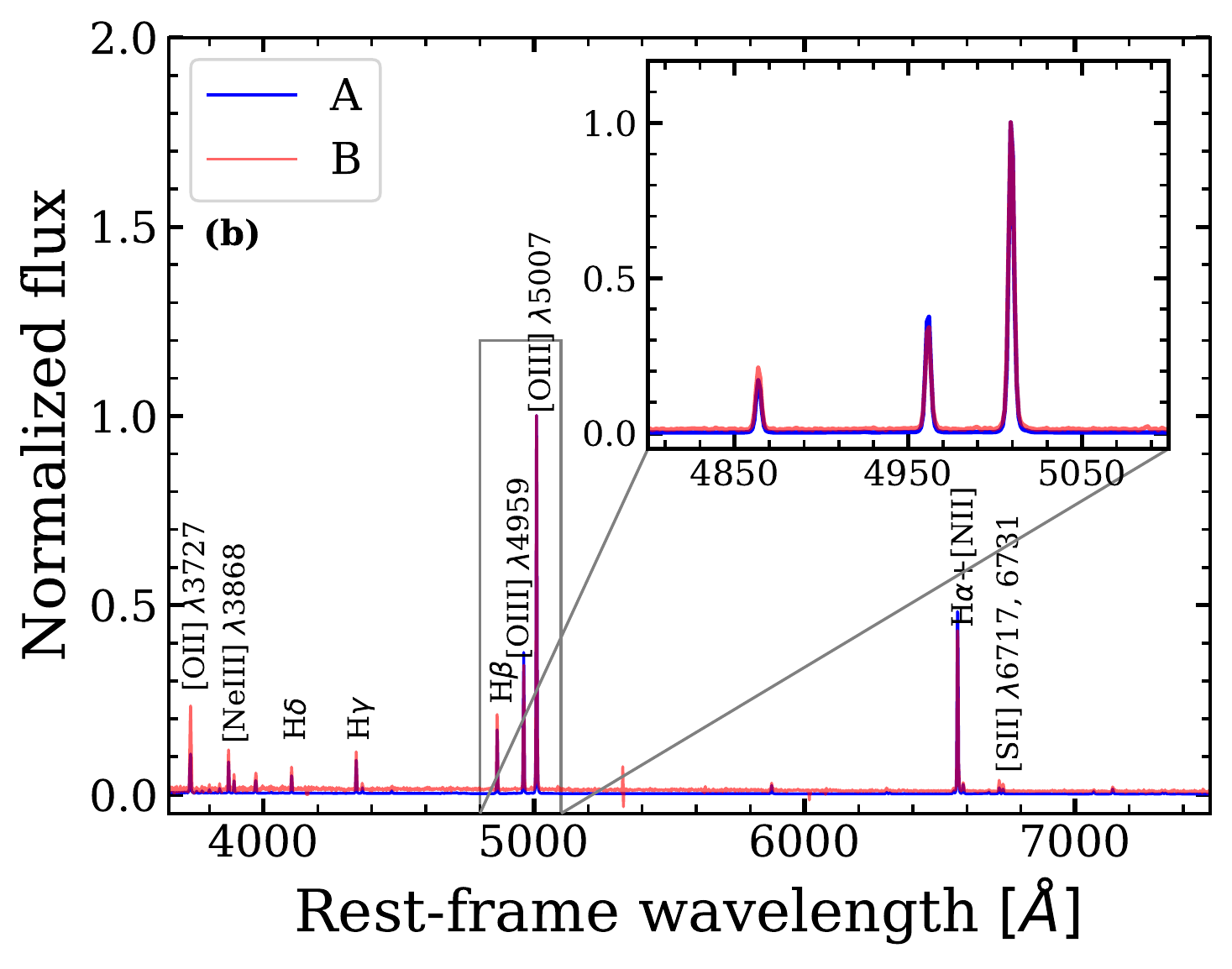}}
\rotatebox{0}{\includegraphics[width=0.5\textwidth]{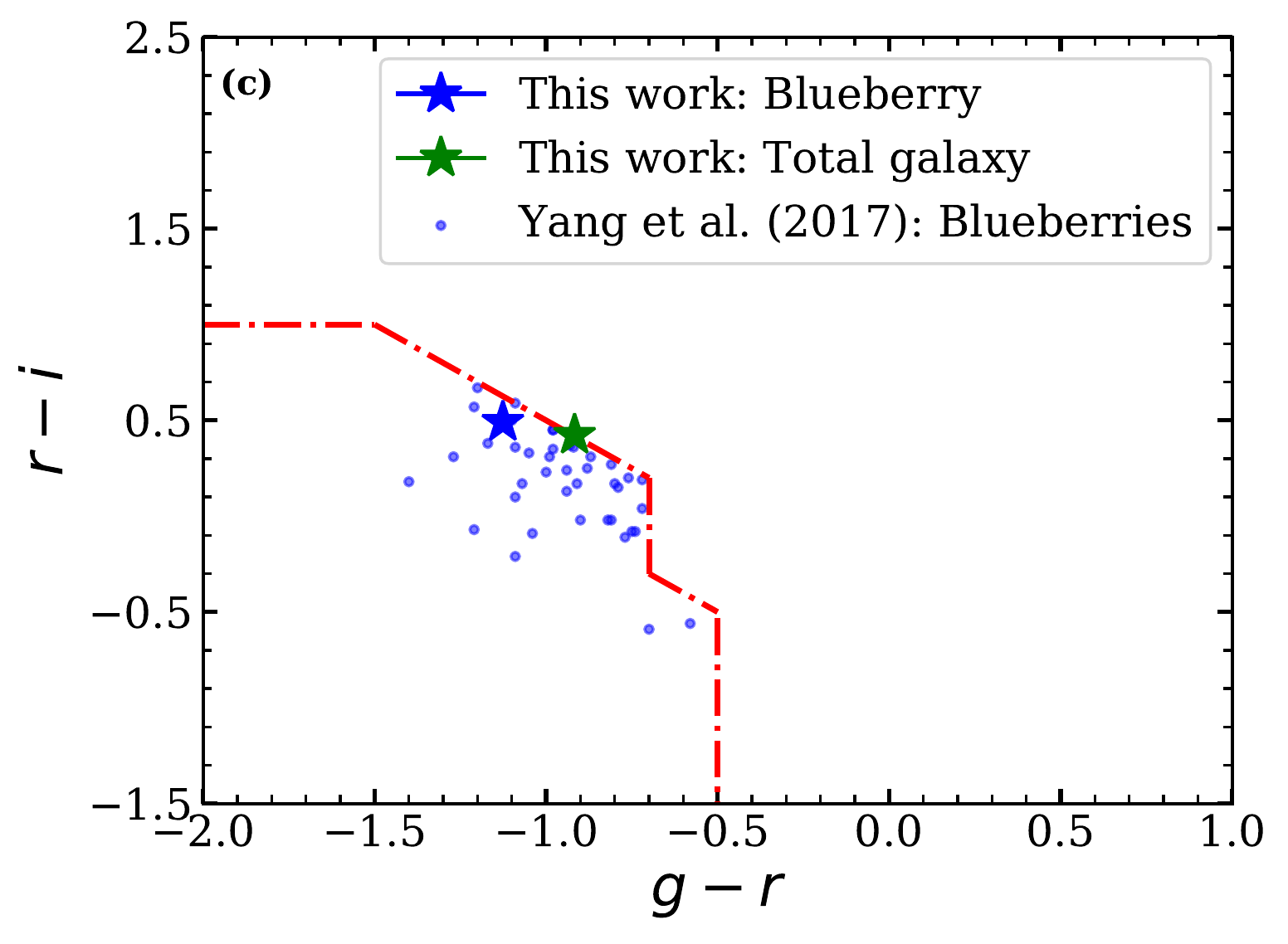}}
\rotatebox{0}{\includegraphics[width=0.45\textwidth]{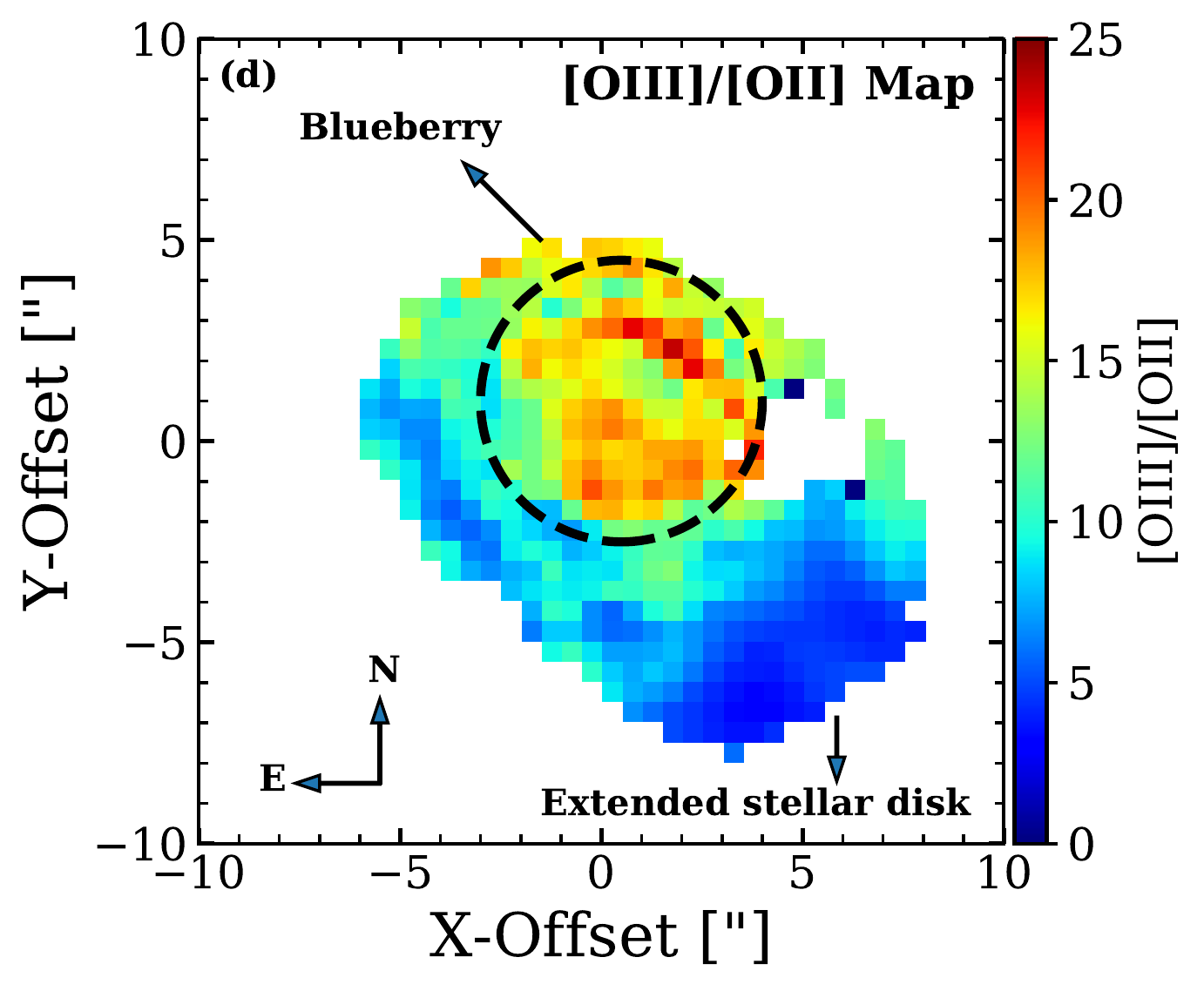}}
\caption{(a) $SDSS$ $i$-band image of the galaxy shown in grey scale overlaid with contours at 3, 6, 12, 24$\sigma$ significance level. The blueberry and extended disk regions are denoted by letters "A" and "B" respectively. (b) One-dimensional spectra extracted from the regions "A" and "B", indicating that these two spectra belong to the same system at z $\sim$ 0.0472. The identified emission line are labeled by their respective names. (c) $g - r$ vs. $r - i$ color-color diagram as used for classification of the blueberries \citet{Yangetal2017}. (d) O$_{32}$ map - the blueberry region is marked by the dashed circle.}
\label{fig:1}
\end{center}
\end{figure*}

In this section, we outline the basic properties of the galaxy SHOC 579. Fig.~\ref{fig:1}(a) displays the SDSS $i$-band image of the galaxy that hosts a bright central compact structure along with surrounding faint disk-like structure within the 3$\sigma$ outer contour. In order to confirm whether both the structures belong to the same system, we extract their corresponding spectra as shown in Fig.~\ref{fig:1}(b). The spatial location of these spectra are marked by A and B in Fig.~\ref{fig:1}(a). In both cases, measurement of redshift based on the strong emission lines e.g., H$\alpha$ and [O III], we confirm that both the compact central structure and the disk-like outer structure are at the same redshift $z$ $\sim$ 0.0472. \\

Fig.~\ref{fig:1}(c) shows the galaxy SHOC 579 on the SDSS $g - r$ vs. $r -i$ color-color diagram. The color-cut marked by the red dash-dot line has been used to classify a galaxy as a blueberry in the low redshift (0.02 $\leqslant$ $z$ $\leqslant$ 0.05) universe \citep{Yangetal2017}. Although the full galaxy lie at the color-cut boundary, the central bright region qualifies as a blueberry at par with other confirmed blueberries identified by \citet{Yangetal2017}. The colors for the central bright region and total galaxy are separately estimated (see Sect.~\ref{psf-match} in the Appendix for details). Although our selected galaxy has been observed earlier in the DR7 of SDSS survey, it is missed out in the previous blueberry sample presented by \citet{Yangetal2017}, most likely due to their selected flag criteria. In their sample, they flagged those galaxies which are represented by "CHILD" in the $SDSS$ survey. Similar to the typical properties of other GPs and blueberry galaxies, the central bright, compact region of our identified blueberry system as marked by the dashed circle in Fig.~~\ref{fig:1}(d) shows a high value of [OIII]/[OII] $\gtrsim$ 10. The values of [OIII]/[OII] drop off sharply as one moves outward, reaching nearly zero at the outskirts of the galaxy, indicating that the outer part is likely dominated by old, low-mass stars.

\subsection{Spatially resolved BPT and other physical properties}
\label{sec:BPT}

\begin{figure*}
\begin{center}
\rotatebox{0}{\includegraphics[width=1.0\textwidth]{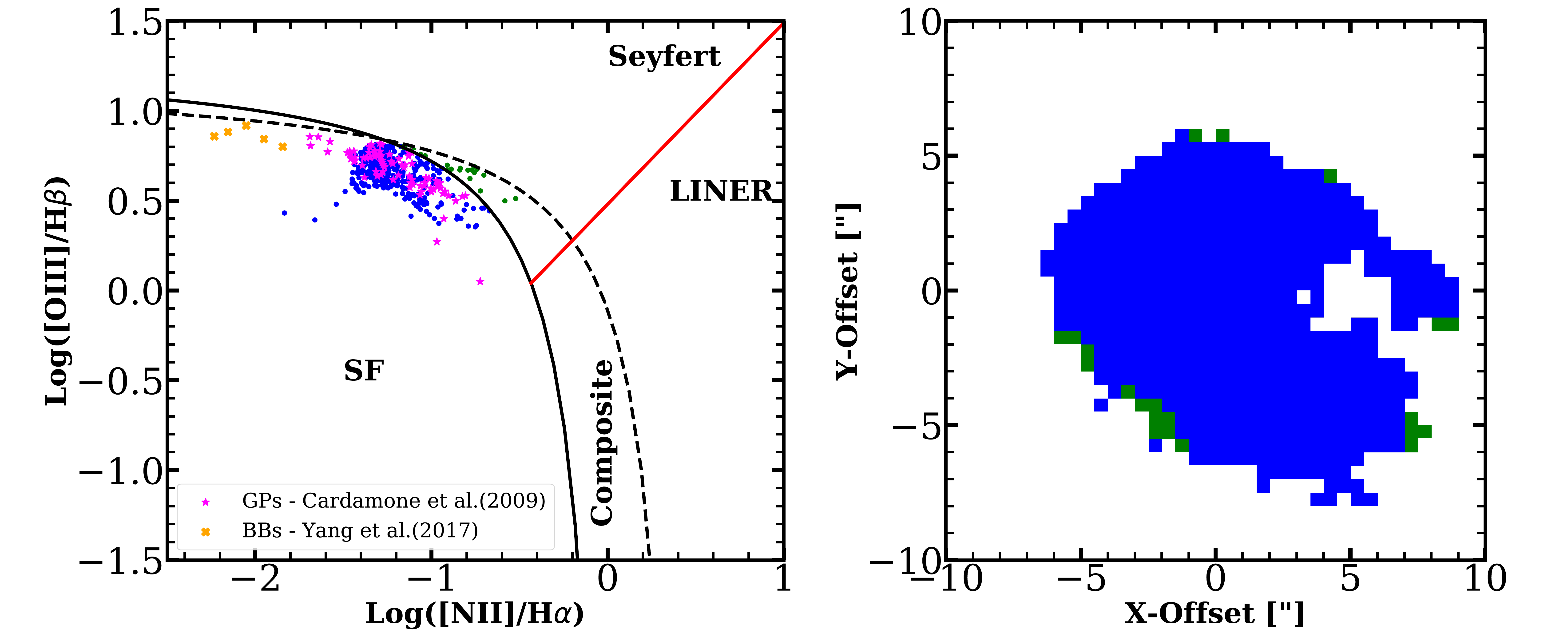}}
\caption{Left: the BPT diagram based on [NII] $\lambda$6583/H$\alpha$ vs. [OIII] $\lambda$5007/H$\beta$ emission line ratios, labeled with different regions corresponding to star formation (SF), AGN/Shocks (Seyfert or LINER) and composite (AGN+SF) processes. The GPs and blueberries taken from the literature\cite{Cardamoneetal2009,Yangetal2017} are shown by star and cross symbols, respectively. The spaxels from our blueberry galaxy are shown by blue dot points. Right: corresponding to the left figure, the color coded 2D spatially-resolved BPT diagram of our blueberry galaxy.}
\label{fig:BPT}
\end{center}
\end{figure*}

In this section, we first derive the spatially resolved Baldwin, Phillips \& Terlevich \citep[BPT;][]{BaldwinPhillipsTerlevich1981} diagnostic diagram to understand the nature of the galaxy spaxel-wise. Spaxels are selected only when their emission lines such as [O III], H$\alpha$, [N II] and H$\beta$ have S/N $\geq$ 3. The BPT diagram (shown in Fig.~\ref{fig:BPT}) indicates that most spaxels either from the blueberry or the LSB disk region fall in the star-formation region - confirming their nature as a star-forming galaxy. Nearly all the spaxels belonging to the blueberry and LSB disk also overlap with the region occupied by other GPs and blueberries available in the literature. This, in other words, confirm that SHOC 579 is not a host to an active galactic nucleus. \\ 

\begin{figure}
\centering
\rotatebox{0}{\includegraphics[width=0.45\textwidth]{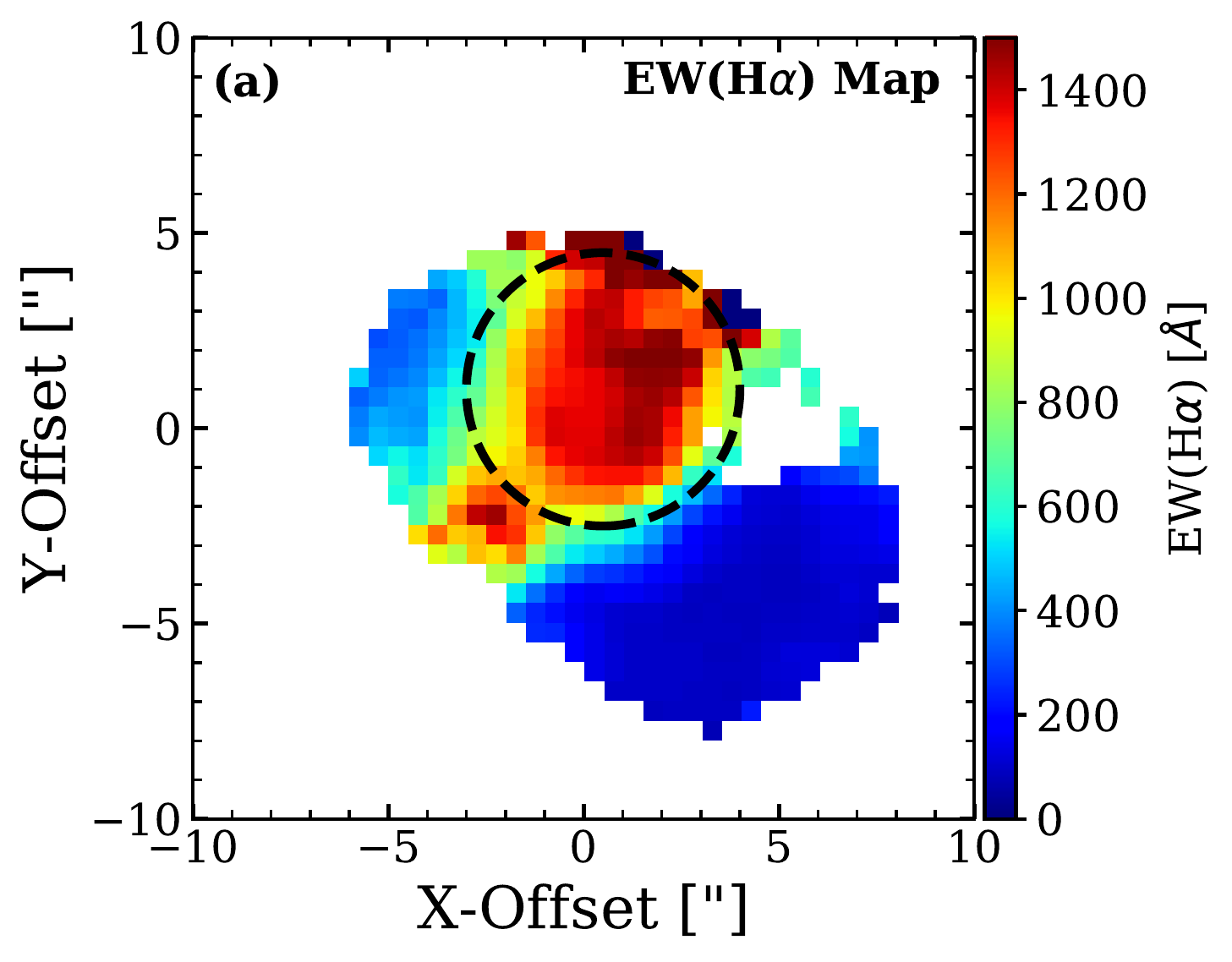}}
\rotatebox{0}{\includegraphics[width=0.45\textwidth]{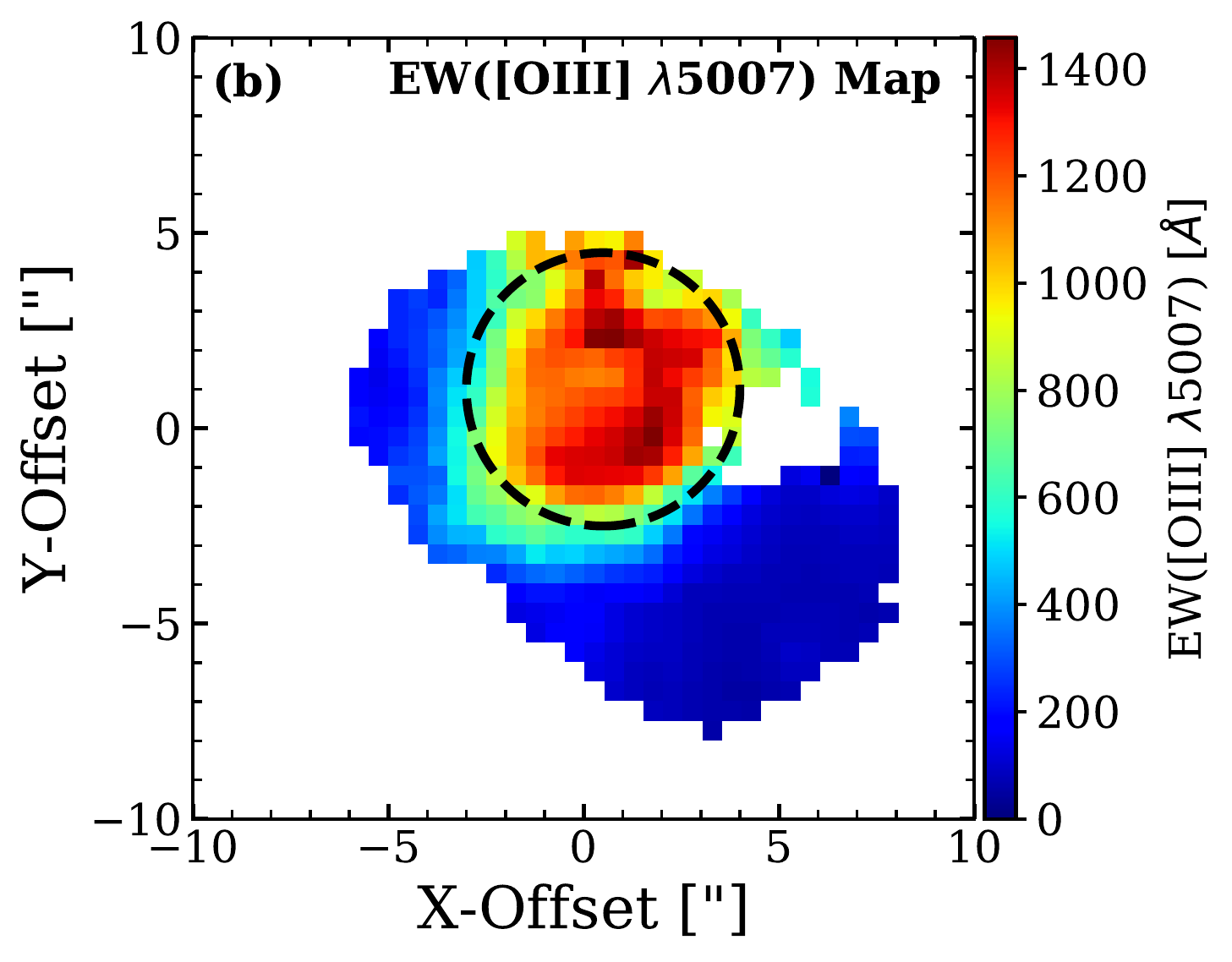}}
\rotatebox{0}{\includegraphics[width=0.45\textwidth]{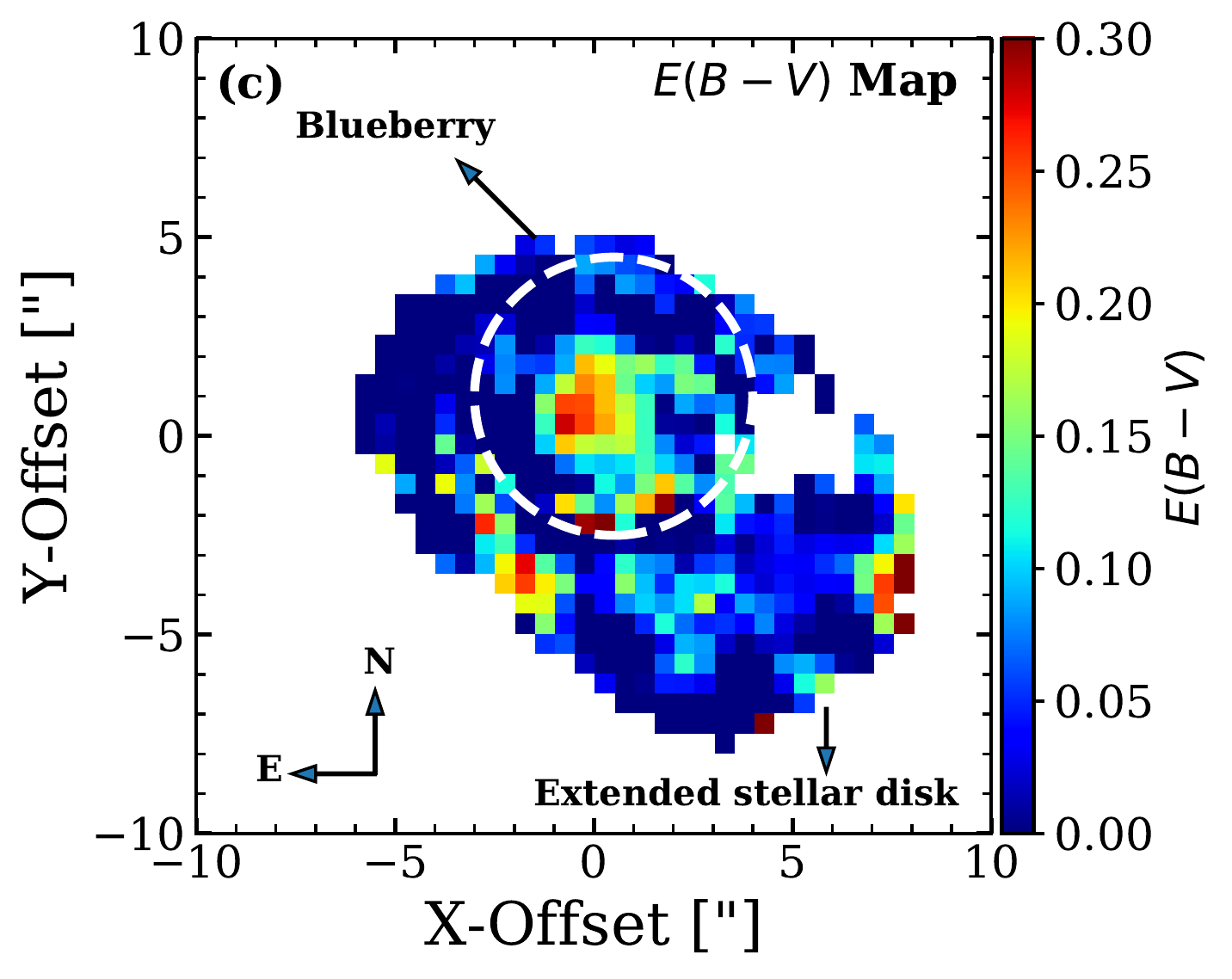}}
\rotatebox{0}{\includegraphics[width=0.45\textwidth]{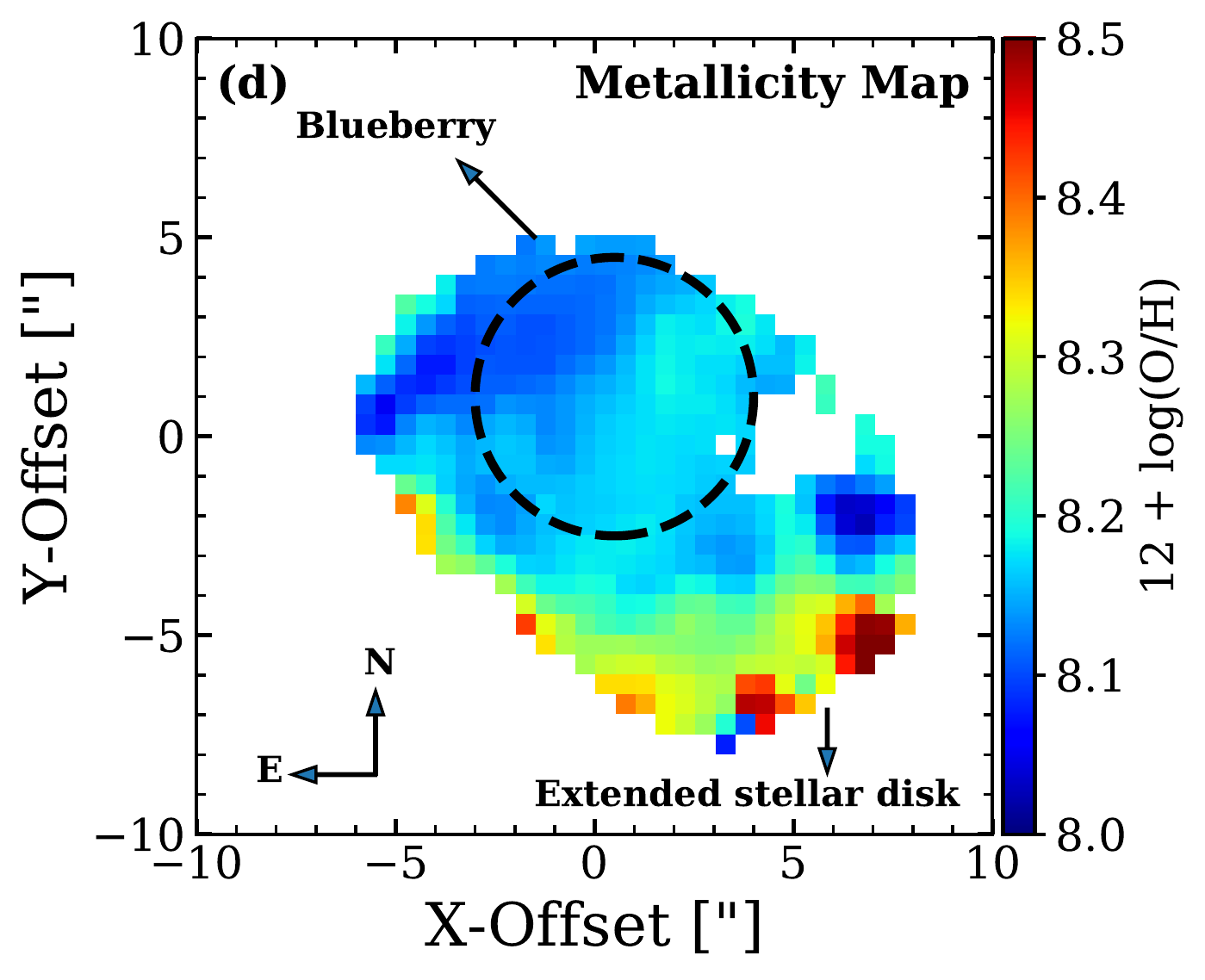}}
\caption{The spatially resolved 2D-maps of (a) EW(H$\alpha$), (b) EW([OIII] $\lambda$5007), (c) $E(B - V)$ and (d) 12+log(O/H) of our galaxy in this work. The dashed circle in each figure indicates the spatial extent of blueberry region in the galaxy.}
\label{fig:maps}
\end{figure}

Since blueberries and GPs are, in general, very compact, their spatially resolved study using the SDSS data alone is not possible. Thanks to the MaNGA IFU data and a close proximity of the blueberry galaxy under this study that allowed us to analyse the spatially resolved spectroscopic properties for the first time. We construct the 2D spatial distribution of EW(H$\alpha$), EW([OIII] $\lambda$5007), E($B - V$) and 12 + log(O/H) parameters using the MaNGA maps as shown in Fig.~\ref{fig:maps} (a) $-$ (d). In these figures, the blueberry region is marked by the dashed circle. The EWs of H$\alpha$ and [OIII] $\lambda$5007 emission lines are directly derived from each spatial spaxel in the MaNGA data cube as discussed in Sec.~\ref{data}. The extremely high values ($> 1200$\AA~ in observed frame or $>1146$\AA~ in rest-frame) of EWs of [O III] and H$\alpha$ in the blueberry region suggest dominance of a very young stellar population over the older ones and an ongoing starburst activity. The rest of the galaxy i.e., the disk-like appears to be like a normal star-forming galaxy. While the central region of the blueberry component is dusty showing $E(B-V) \simeq 0.2$ or above, the rest of the galaxy is almost dust-free having $E(B-V) < 0.1$.\\

The metallicity map (shown in Fig.~\ref{fig:maps}d) is derived using the so-called N2-method \citep{PettiniPagel2004},

\begin{equation}
12 + \log(O/H) = 8.90 + 0.57 \times \log([NII] \lambda 6583/H\alpha).
\end{equation}

\noindent The metallicity map clearly shows that the blueberry region has a sub-solar metallicity. The North-East corner is slightly metal poor compared to the rest. Note that this is also the region having least amount of dust (as seen in the $E(B-V)$ map). The rest of the galaxy is also at sub-solar metallicity, except the South edges of the galaxy that show a relatively metal-rich regions. 

\subsection{Stellar mass and Star-formation Main Sequence}
\label{sec:SFR}

\begin{figure*}
\begin{center}
\rotatebox{0}{\includegraphics[width=1.05\textwidth]{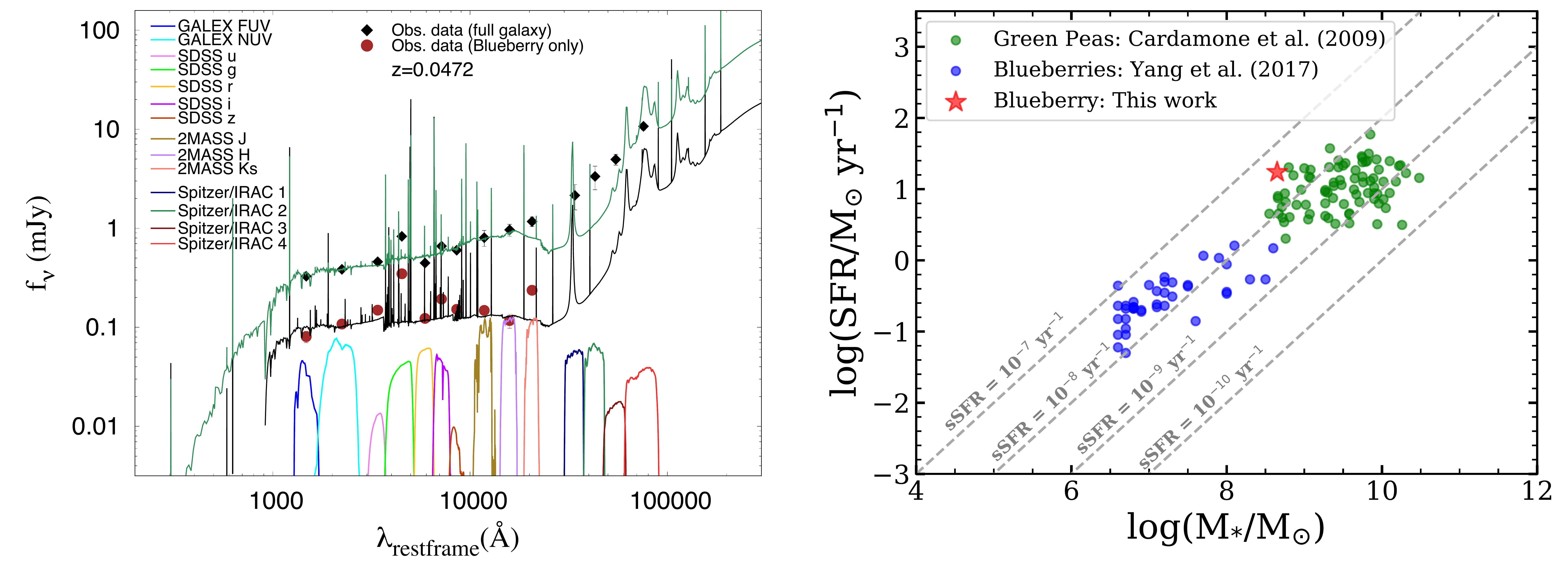}}
\caption{Left panel: The multi-wavelength SED modelling $-$ the black diamond and red circle points represent the photometric band fluxes from FUV to IR for full galaxy and blueberry component only, respectively, obtained using PSF matched images (see Sect.~\ref{psf-match}). The corresponding SED spectra for the full galaxy and blueberry component are shown by green and black solid lines, respectively. The error bars denote $1\sigma$ uncertainties on the measured fluxes.
Right panel: The main-sequence relation for blueberries and GPs. The blueberry in this work is shown by star symbol. Other blueberries and GPs studied in the literature \citep{Cardamoneetal2009,Yangetal2017} are represented by blue and green solid circles, respectively. Several dashed-straight lines show different MS relation at constant sSFR from 10$^{-7}$ to 10$^{-10}$ yr$^{-1}$.}
\label{fig:MS}
\end{center}
\end{figure*}

Here we characterize the stellar properties of the blueberry and its extended host galaxy by modelling their spectral energy distribution (SED). A multi-wavelength broad-band SED ($1300 - 45000$ \AA) is constructed by estimating the photometric fluxes from far-ultraviolet (FUV) from GALEX \citep{Martin2005} to infrared (IR) from Spitzer survey \citep{Dale2009}. Since the multi-band observations are taken with telescopes having widely varying point spread function (PSF), we employ PSF-matching technique \citep{Becker2012} to estimate the fluxes from the central blueberry region and the full galaxy (see Sec.~\ref{psf-match} in the Appendix for details). The resulting SED is modelled with the Python Code Investigating GALaxy Emission \citep[PCIGALE;][]{Boquienetal2019} code (details are provided in Sec.~\ref{sed} in Appendix). The best-fit SED model, presented in the left panel of Fig.~\ref{fig:MS}, has a $\chi_{reduced}^2=8.0$ and $5.2$ for the full-galaxy model and blueberry component respectively. Note that our full galaxy model deviates considerably from the data on the high wavelength side, especially on two Spitzer bands. Our best-fit model SED yields a total stellar mass of $5.2\times 10^{9}$~M$_{\odot}$ and $4.5\times 10^{8}$~M$_{\odot}$ for the full galaxy and the blueberry component, respectively. Note that the stellar mass of the blueberry component is only about 8\% of the full galaxy, consistent with that observed in the case of GPs \citep[e.g.,][]{Amorinetal2012}. This is the most massive blueberry known to date similar to GPs \citep[][also see right panel of Fig.~\ref{fig:MS}]{Cardamoneetal2009}, although this is as compact (with $R_e \sim 217$~pc as estimated in Sec.~\ref{sec:disk}) as other locally known blueberries.\\

Based on our SED modelling, we find that the SFR, averaged over 10 Myr of the SFH, of the full galaxy is $\sim 15.1$~M$_{\odot}$~yr$^{-1}$. 
Whereas, the extinction corrected H$\alpha$ SFR in the blueberry region is found to be $\sim 17.4$~M$_{\odot}$~yr$^{-1}$ - this is the highest star-forming blueberry amongst the local blueberries known so far. The mass-weighted age of the stellar population in the Blueberry component and the full galaxy are $5.6$~Gyr and $7.05$~Gyr respectively. Our SED age estimates are comparable to that obtained based on the pPXF fitting of the stacked spectra from either regions (see Appendix~\ref{sed} and~\ref{int-spat-ana} for further details).
The right panel of Fig.~\ref{fig:MS}, shows the star-formation main-sequence (MS; M$_{*} -$ SFR) relation for our blueberry galaxy and a comparison with other blueberries and GPs in the literature \citet{Cardamoneetal2009,Yangetal2017}. The stellar mass of the blueberry in this work is obtained from our SED analysis (see Sec.~\ref{sed}), while the SFR is derived using the dereddened H$\alpha$ luminosity from each spaxel of the galaxy, following the relation \citep{Kennicutt98}: 
 \begin{equation}
 SFR(M_{\odot}~yr^{-1}) = 7.9 \times 10^{-42} L_{H\alpha}(erg~s^{-1})
 \end{equation}

\noindent The estimated SFRs from each spaxel are then integrated to estimate the total SFR. On the right panel of Fig.~\ref{fig:MS}, we see that blueberries and GPs from the literature are higher by $\sim 2 -3$ orders of magnitude in terms of sSFR compared to normal star-forming galaxies (represented by the constant sSFR line 10$^{-10}$~yr$^{-1}$); our blueberry galaxy is not exceptional to this. This implies that the typical mass doubling time for GPs and blueberries is between 100 Myr and $\sim$ 1 Gyr. The blueberry galaxy SHOC 579 studied in this work is consistent with these time scales.

\subsection{Comparison with other GPs and Blueberries}
\label{sec:comparison}

\begin{figure*}
\begin{center}
\rotatebox{0}{\includegraphics[width=0.4\textwidth]{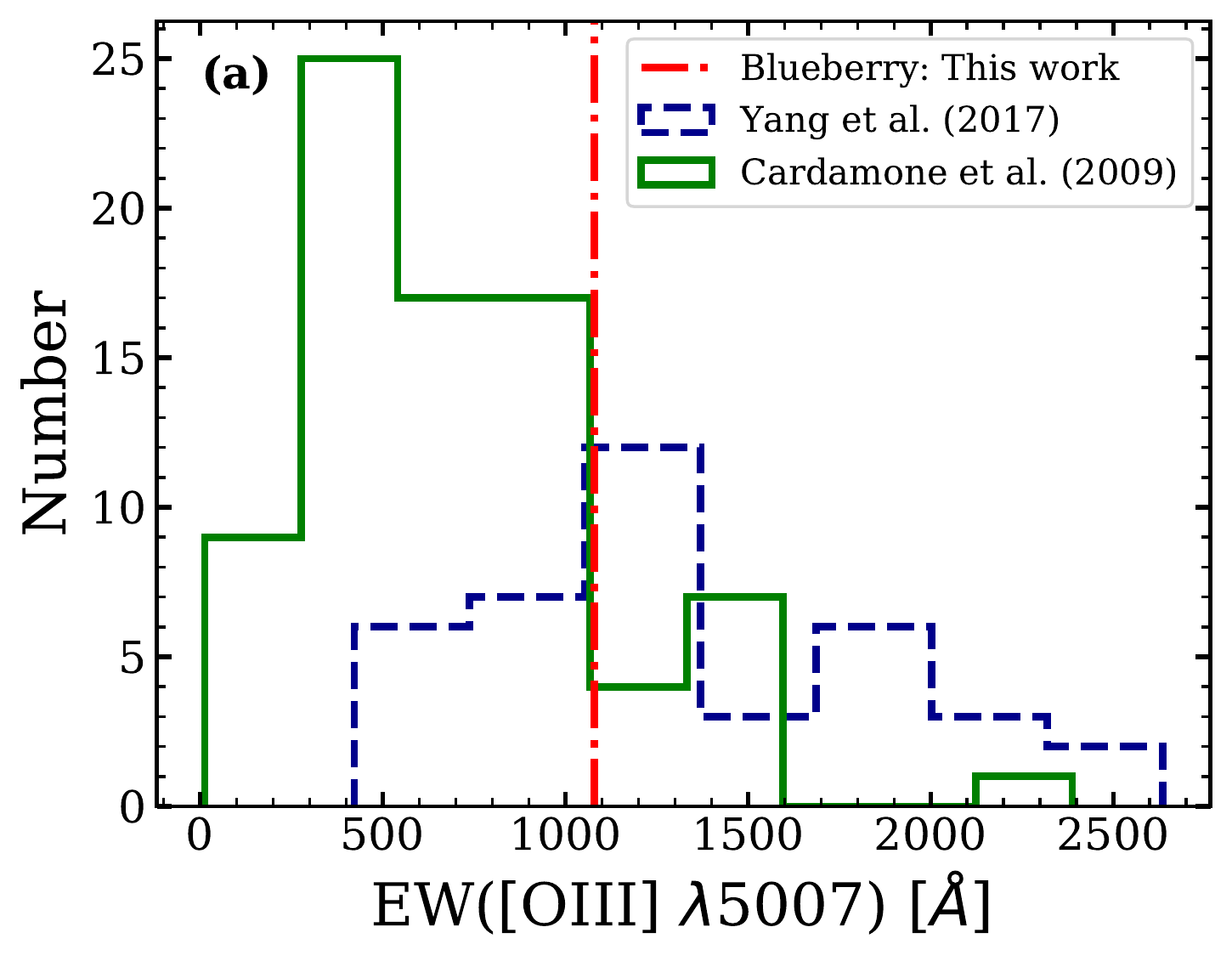}}
\rotatebox{0}{\includegraphics[width=0.4\textwidth]{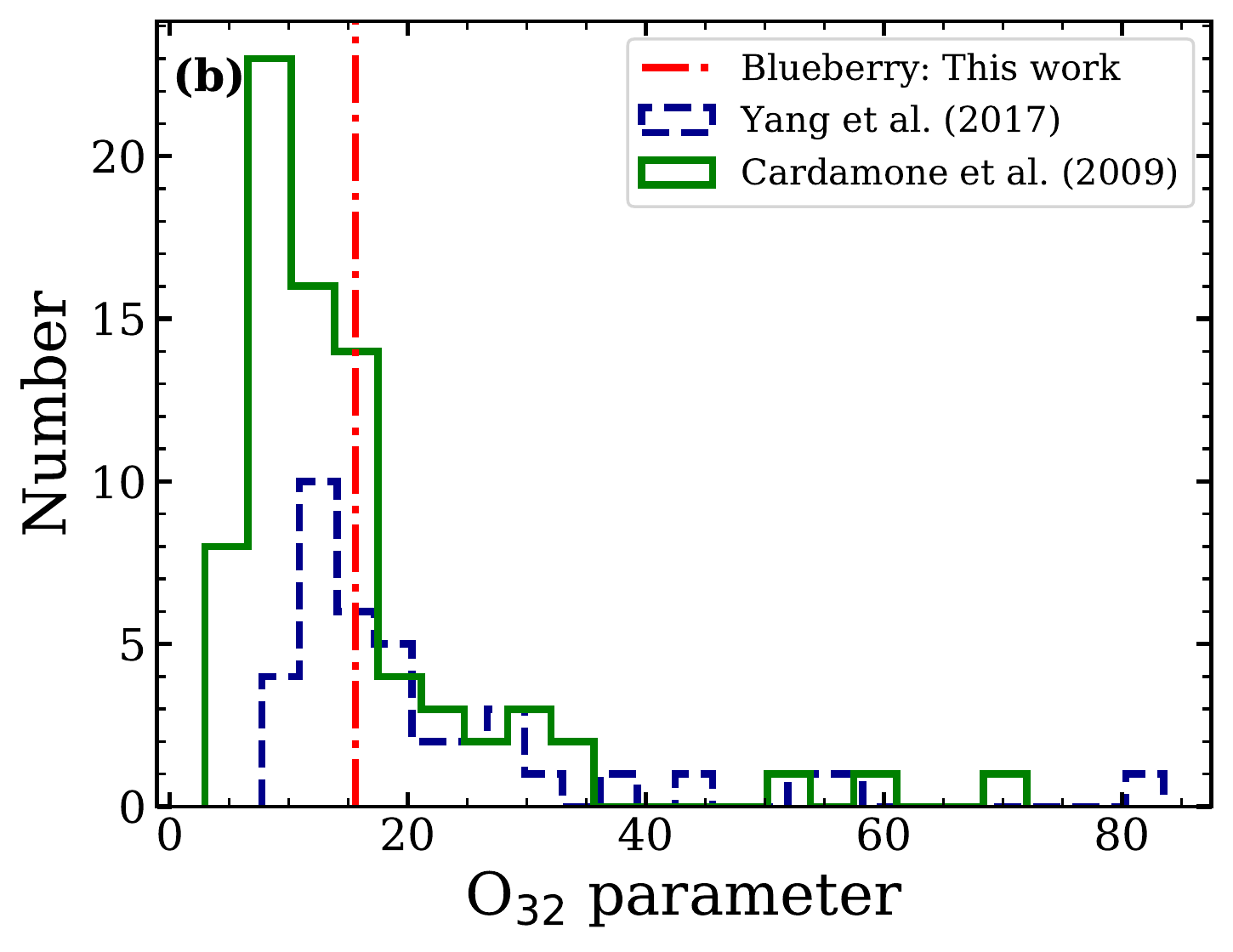}}
\rotatebox{0}{\includegraphics[width=0.4\textwidth]{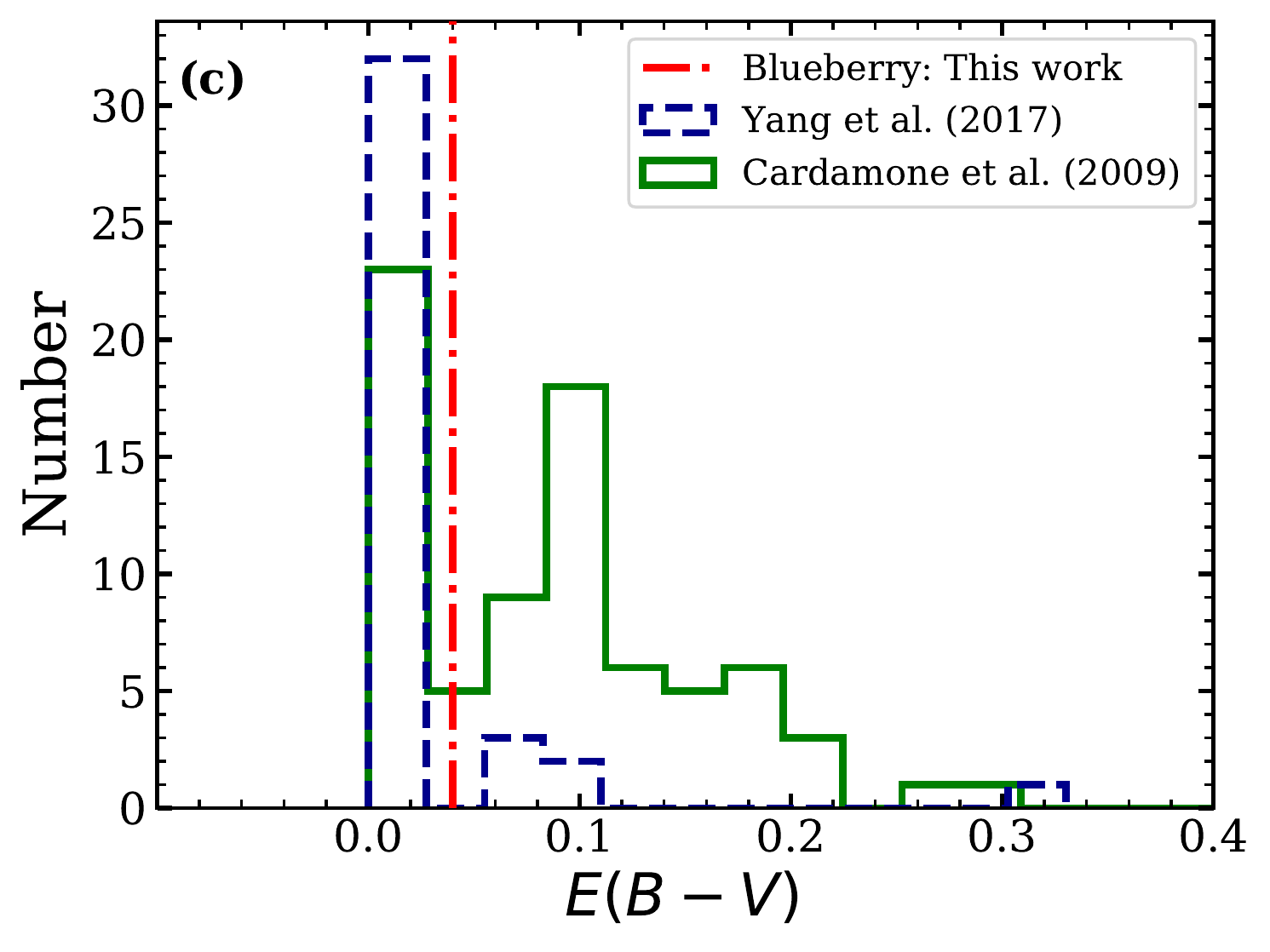}}
\rotatebox{0}{\includegraphics[width=0.4\textwidth]{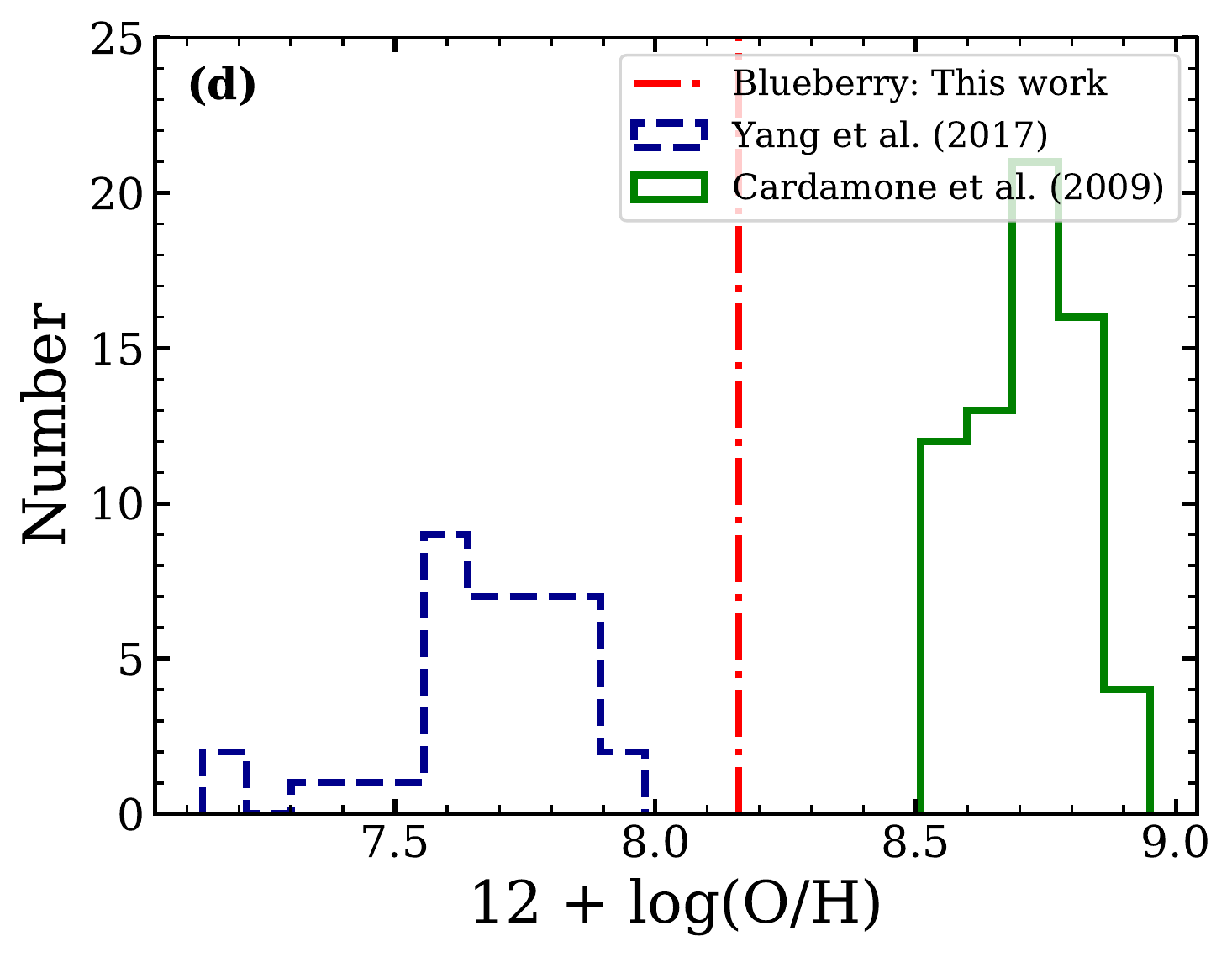}}
\caption{The histograms of (a) EW([OIII] $\lambda$5007), (b) O$_{32}$ parameter, (c) $E(B - V)$ and (d) 12+log(O/H) of GPs and blueberries studied in the literature \citep{Cardamoneetal2009,Yangetal2017}, respectively, represented by green solid and blue dashed lines. The median values of the same parameters estimated over the blueberry region of SHOC 579 are shown by vertical red dot-dashed line.}
\label{fig:histogram_compare}
\end{center}
\end{figure*}

Fig.~\ref{fig:histogram_compare} (a) $-$ (d) show the comparison of various physical parameters such as EW [OIII] $\lambda$5007, [OIII]/[OII] ratio, nebular color excess, metallicity of the blueberry component (i.e., within the dashed circle) of SHOC 579 with previously known GP and blueberry galaxies available in the literature \citep[e.g.,][]{Cardamoneetal2009,Yangetal2017}. While these parameters for the GPs and blueberries taken from the literature are derived using the aperture integrated light obtained with slit observations (i.e., the SDSS and MMT spectroscopic observations), the same for our blueberry component is represented by the median values of all the spaxels within dashed circle as shown in Fig.~\ref{fig:maps}.
Our obtained median values of EW [OIII] $\lambda$5007, [OIII]/[OII] ratio, and $E(B-V)$ are similar to other typical GPs and blueberries, except the metallicity. The median metallicity of our blueberry component has an intermediate value, laying between the typical blueberries and GPs. In other words, our blueberry is metal-poor compared to GPs and metal-richer compared to blueberries. In terms of stellar mass, our blueberry is similar to typical GPs (see right panel of Fig.~\ref{fig:MS}). Overall, our comparison implies that the galaxy under this study represents the most metal-rich and massive blueberry source. \\

\section{2D modelling of the surface brightness distribution}
\label{sec:disk}

\begin{figure}[!hhhhhhhhhhhhhh]
\begin{center}
\rotatebox{0}{\includegraphics[trim=0cm 0cm 0cm 0cm, clip=true, width=0.7\textwidth]{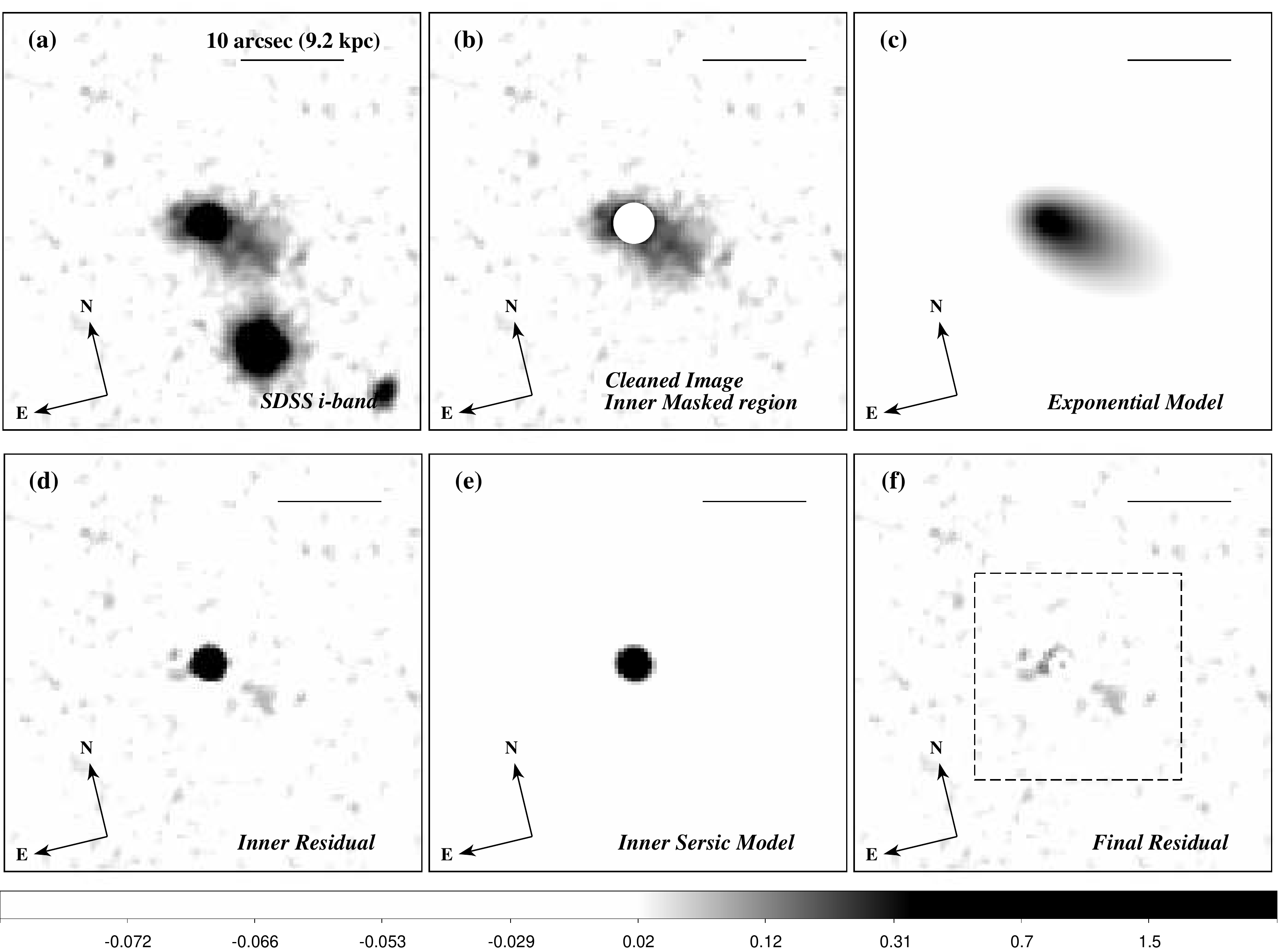}}
\caption{(a) The observed SDSS $i$-band image of the target galaxy. (b) The cleaned (nearby sources removed) parent image of the source used for 2D fitting. Here, the inner central region has been masked with a circular aperture of radius $\sim$2 arcsec while fitting the 2D exponential disk (for details see text). (c) The modelled lopsided exponential disk. (d) The central bright component of the galaxy obtained as residual, after subtracting the modelled exponential disk from the observed galaxy. (e) The modelled Sersic component to the inner/central region of the galaxy. (f) The final residual image after subtracting the complete galaxy model from the observed galaxy. The dashed box here represents the region occupied by the source in the parent image. In each image, a common scale limit and contrast bias are used. The colorbar units is in nanomaggy. }
\label{galfit}
\end{center}
\end{figure}

We carry out a detailed 2D modelling of the galaxy using GALFIT \citep{Pengetal2002}. In that, we model the central bright component using a Sersic profile \citep{Sersic1968} while the extended faint stellar envelope with an exponential profile. The best-fit models and residues are displayed in Fig.~\ref{galfit}. The details of our GALFIT modelling is presented in Sect.~\ref{Galfit} (see Appendix). Based on the two-component modelling of the SDSS $i$-band image of the galaxy, we arrive at an insightful revealing about the nature of the galaxy's light distribution. The best-fit model of the extended host galaxy reveals a faint, lopsided, lower-surface brightness exponential profile that is presumably due to an old stellar host. The presence of an exponential profile as found in many BCDs \citep[e.g.,][]{Lianetal2015, Papaderos2002} as well as in SHOC 579, is not convincing enough for a thin stellar disk; possibilities of a triaxial stellar system cannot be ignored. The central surface brightness of the disk is $\mu_{0}=22$~mag~arcsec$^{-2}$ \citep[e.g.,][]{Brown2001,Adami2006,Schombertetal2001,Pahwa2018} with a scale-length of $R_d=1.54$~kpc. Note that the disk is relatively smaller in size and the scale length is similar to that of dwarf LSB galaxies \citep{Schombert1995,Schombertetal2001,Papaderos1996b,GildePaz2005}. The Sersic profile for the inner component has a Sersic index of $n=1.48$ and effective radius of $217$~pc. Since our GALFIT modeling is performed using the SDSS $i$-band image, hence it does not contain strong nebular emission lines (e.g., H$\alpha$ line) for the given redshift of the galaxy. Nevertheless, the contributions from nebular continuum cannot be ruled out. In fact, the contribution from nebular continuum has especially been found significant in the SDSS $i$-band image \citep[e.g.,][]{Izotov2011}. Therefore, an exponential light distribution indicating a stellar disk-like structure around blueberry source may be a generic property of extended nebular halo, similar to that seen in several luminous BCDs \citep[e.g.,][]{Papaderos2002}. \\

\begin{figure*}[!hhhhhhhhhhhhhhhhhh]
\begin{center}
\rotatebox{0}{\includegraphics[width=0.49\textwidth]{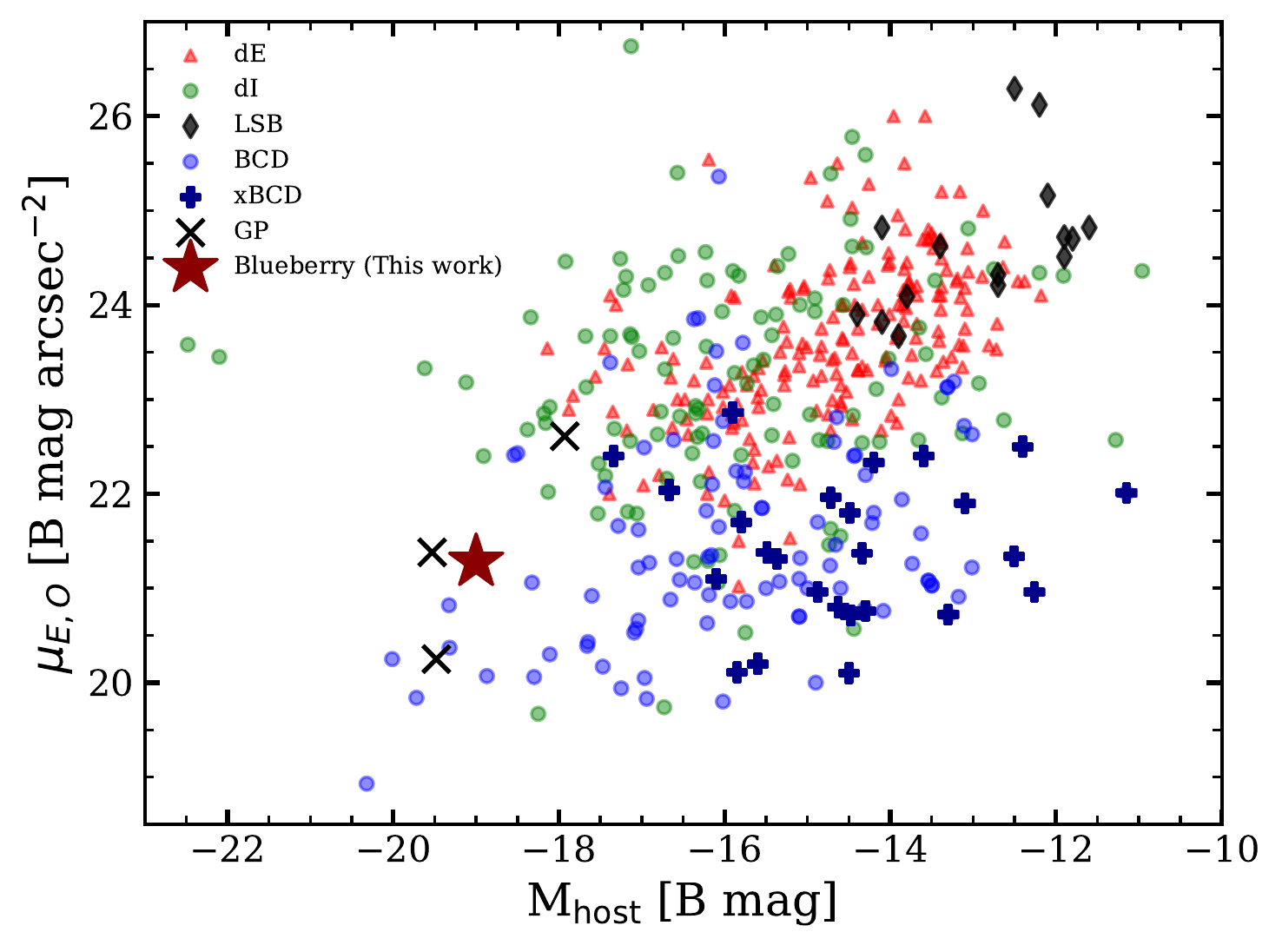}}
\rotatebox{0}{\includegraphics[width=0.49\textwidth]{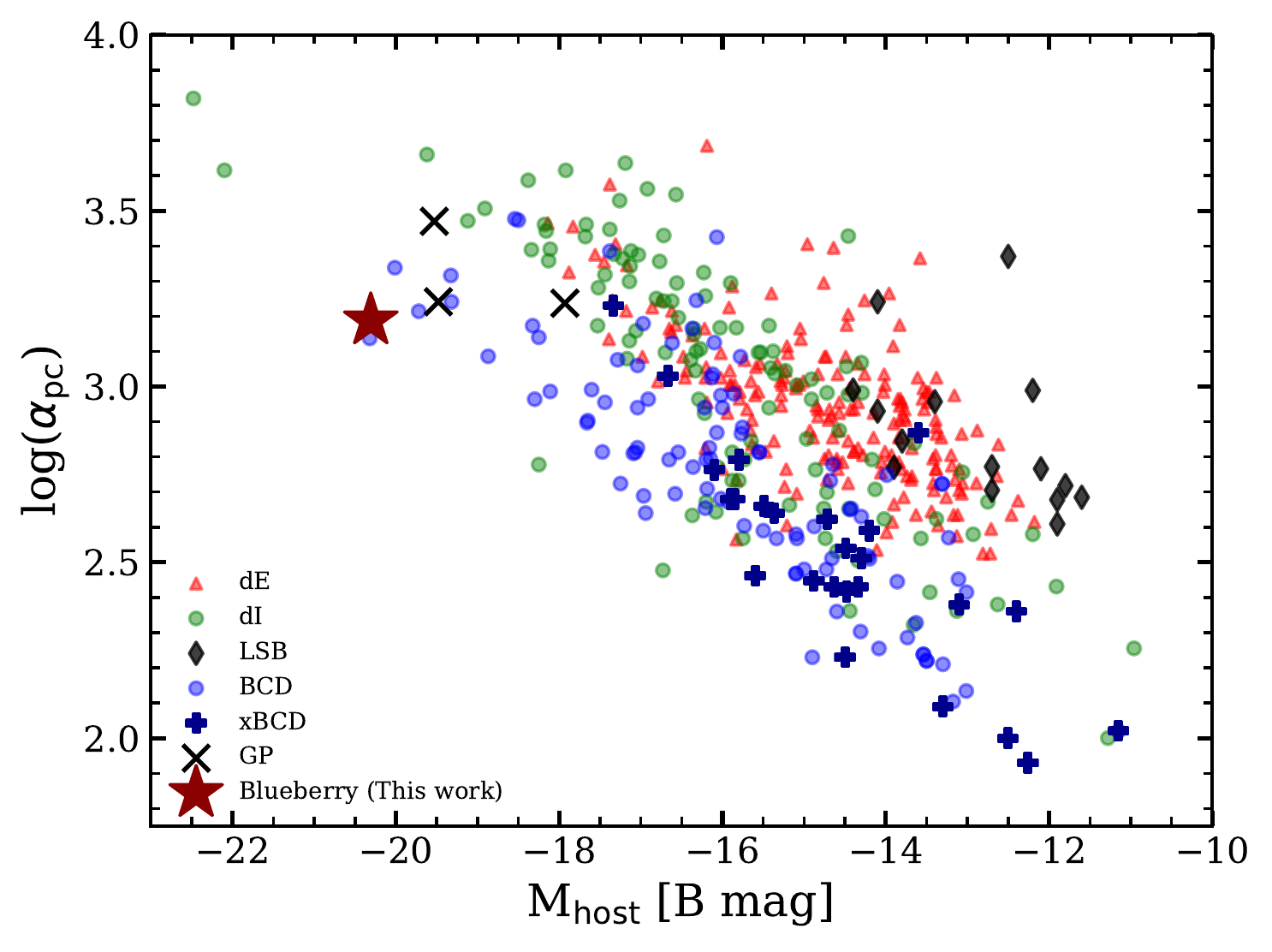}}
\caption{Comparison of structural properties of the host LSB galaxies of blueberry (as shown by star symbol) with other class of low-mass dwarf galaxies such as dwarf elliptical (dE), dwarf irregular (dI), BCD, XBCD and LSB taken from Papaderos et al. (2008), including three GPs from \citet{Amorinetal2012}. Left and right panels compare, respectively, the central surface brightness and the exponential scale length vs. the absolute magnitude of the LSB host galaxy of different class of dwarf systems. The respective symbols for representing the galaxy of different types are shown in the legends.}
\label{fig:structure_compare}
\end{center}
\end{figure*}

We compare the structural properties of the blueberry host galaxy with those of local dwarf galaxies, including LSB, GP and XBCD systems (see Fig.~\ref{fig:structure_compare}). Since the image of SHOC579 is not available in the Johnson $B$-band, the values of central surface brightness ($\mu_{E,O}$) and absolute magnitude (M$_{B}$) of our blueberry host in the $B$-band are therefore derived using the SDSS images in $g$ and $r$-band, after applying the Lupton transformation relations published on the SDSS DR4 website. Note that the derived values of $\mu_{E,O}$ and M$_{B}$ have been corrected for Galactic extinction. In Fig.~\ref{fig:structure_compare}, it can be clearly seen that our blueberry along with GPs fall in the same parameter space that is populated by luminous BCDs. This result suggests that all these galaxy classes have common structural properties, and an extended disk-like structure in starburst dwarf galaxies such as blueberries, GPs and XBCDs might be a general rule. Alternatively, this result also implies that, with respect to
the structural properties of their host galaxies, blueberries and GPs pass through a common evolutionary track along with the main population of BCDs, except their extreme emission line properties observed due to their strong starburst phase. Their strong starburst phase is likely due to an occasional event whose triggering mechanism needs to be explored in detail.\\

In the context of above discussion, it is worth mentioning that the possibility of a very faint old stellar disk in blueberry host cannot be dismissed. A detailed analysis of structural properties of CELLs (i.e., blueberries and GPs) host galaxies using near-infrared (NIR) images having a spatial resolution of sub-kpc scale (e.g., observations of CELLs galaxies with JWST) may further shed light on the existence of old stellar disk. Such studies with the NIR images could also reveal additional stellar masses distributed over larger radii.

\section{Evidence of old stellar population}
\label{old-pop}

\begin{figure*}
\begin{center}
\rotatebox{0}{\includegraphics[width=1.\textwidth]{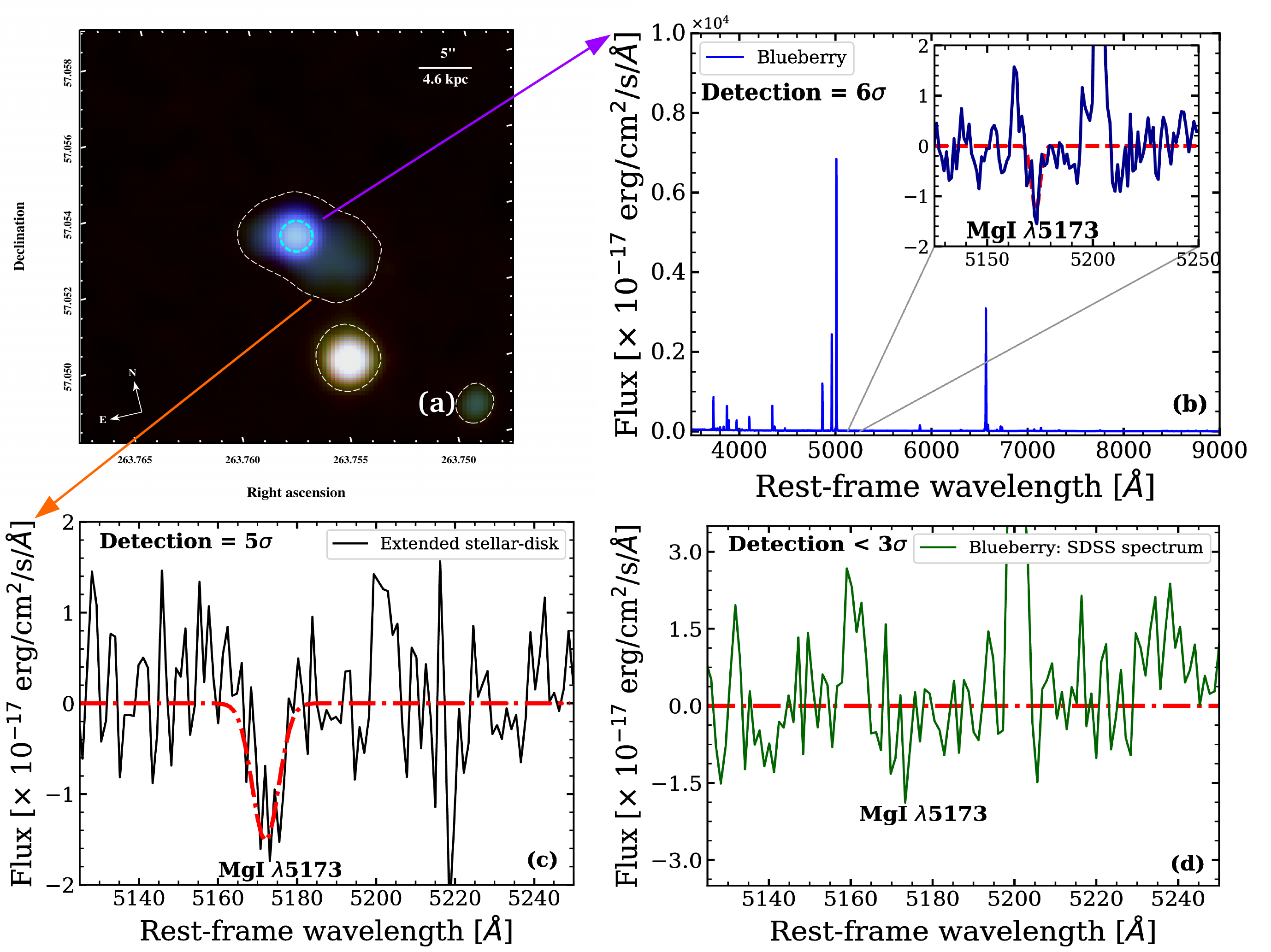}}
\caption{(a) The \textit{gri-band} color composite image of SHOC 579. The dashed line shows contour at 3$\sigma$ level above the background. Inner dashed circle of size 3'' (diameter) marks the blueberry region. (b) The stacked full observed spectrum within 3" diameter. The inset represents the detection of MgI $\lambda$5173 absorption line from the blueberry region. (c) The detection of MgI $\lambda$5173 absorption line obtained through the stacked spectra from extended diffuse stellar disk region. (d) Null detection of MgI $\lambda$5173 absorption line in the observed SDSS spectrum taken over central 3" diameter on the blueberry region. In each spectrum, the red dashed line represents fitted continuum + Gaussian models over the observed MgI$\lambda$5173 absorption feature. In panel d, the Gaussian model is not fitted as the detection of MgI$\lambda$5173 absorption feature is below 3$\sigma$.}
\label{fig:MgI}
\end{center}
\end{figure*}

Fig.~\ref{fig:MgI} presents a clear detection of Mg{\sc i} $\lambda$5173,5167 absorption line in the central blueberry region as well as in the LSB disk. The Mg{\sc i} $\lambda$5173,5167 absorption line is generally seen in the late-type stars, and is an excellent tracer of $\alpha$-abundances in a galaxy. Its detection in the galaxy's spectrum provides us the confirmation of the presence of old stars \citep[e.g.,][]{Amorinetal2012}. However, the Mg{\sc i} $\lambda$5173,5167 absorption lines are too weak to be detected in the spatially resolved spaxel in the galaxy. Even in the SDSS 3" fiber spectrum centred on the blueberry region, our search resulted in a weak detection (below 3$\sigma$; see Fig.~\ref{fig:MgI}d). It is only after stacking the MaNGA spectra from both the blueberry and the LSB disk regions (for details, see Sec.~\ref{int-spat-ana} in Appendix), we could reveal the detection with confidence. In Fig.~\ref{fig:MgI}b,c, we show the Mg{\sc i} $\lambda$5173,5167 absorption feature detected with a significance of 6$\sigma$ and 5$\sigma$ in the blueberry and disk region respectively - {\it indicating clear evidence of an underlying old stellar population in the blueberry region}. Thanks to the MaNGA IFU observations \citep{Bundyetal2015} with large integration time $\sim 8100$s that made detection possible. In fact, this is the first time, we observationally establish a clear existence of an old stellar population simultaneously in the blueberry region and in the surrounding LSB disk using IFU spectroscopic data. Future IFU observations of other blueberries and GPs would be an useful path to unravel their true stellar population content. It is worth mentioning here that a very recent study based on resolved structure and color deduced with the newly acquired HST imaging observations has also revealed the evidence of old stellar populations ($\textgreater$ 1 Gyr) in a set of nine GP galaxies \citep{Clarkeetal2021}.\\ 

\begin{figure*}
\begin{center}
\rotatebox{0}{\includegraphics[width=1.\textwidth]{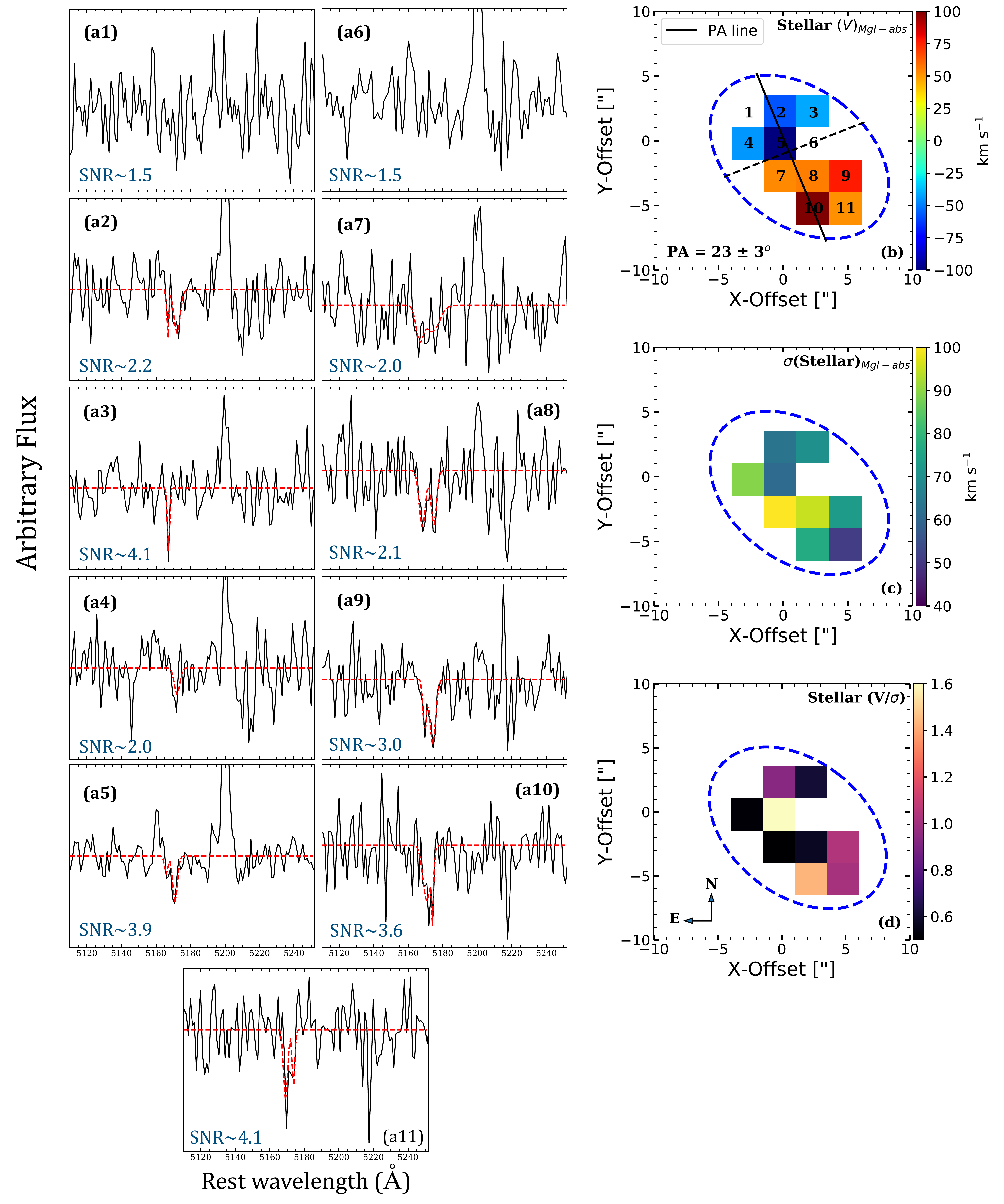}}
\caption{Left panels: a1 - a11 show the spectra for detection of Mg{\sc i} $\lambda$5173,5167 absorption line. Each spectrum is derived after stacking the spectra from all the possible 5 $\times$ 5 MaNGA spaxels over the galaxy extent as denoted by dashed blue ellipse in the right panel. Here each 5 $\times$ 5 binned spaxels is represented by a single spaxel and numbered according to their corresponding 1D spectra (a1 - a11). The Gaussian models (red dashed lines) are only fitted over the 1D spectrum showing the line detection with SNR above 1.5. In the right panel, (b) shows the stellar line-of-site (LOS) rotation velocity ($V$) derived using stellar Mg{\sc i} $\lambda$5173,5167 absorption line. Here solid and dashed lines represent major and minor kinematic rotation axes of stellar component, respectively. Similarly, (c) and (d) show velocity dispersion ($\sigma$) and its $V/\sigma$ maps of stellar component.}
\label{fig:MgI-stellar-kin}
\end{center}
\end{figure*}

We further made our effort to detect the Mg{\sc i} $\lambda$5173,5167 absorption feature in a gradual spatially-resolved manner, after stacking the spectra from 5 $\times$ 5 spaxels over the whole galaxy extent (e.g., see left panel in Fig.~\ref{fig:MgI-stellar-kin}a1-a11). We selected a 5 $\times$ 5 spaxels stacking because it gives us the spectral information over a spatial region of 2.5" that is equivalent to SDSS fiber slit aperture and also to the effective spatial resolution of MaNGA survey (i.e., 2.5" FWHM). In this manner, we find a weak (below 3$\sigma$) detection of Mg{\sc i} $\lambda$5173,5167 absorption, except a few regions in the LSB disk and central blueberry region. Nevertheless, the presence of weak Mg{\sc i} $\lambda$5173,5167 spectral feature through out the galaxy extent suggests that our blueberry host galaxy contains a very faint and wide-spread old stellar population. Whether the stellar disk with old stellar population is rotation supported is addressed in the section~\ref{kin}. Our analysis at this stage, provides a clue that there might exist an underlying old stellar disk in CELLs galaxies, and it further demands a detailed exploration with a large sample of CELLs galaxies with a better quality data. \\

\par
Intriguingly, the metallicity of the blueberry region is only sub-solar which is obtained as $Z \sim 0.008$ from our SED analysis. The SED-based metallicity is consistent with the N2-based metallicity map derived from the MaNGA observation for a solar [O/Fe] ratio (see Fig.~\ref{fig:maps}d), while the disk is at about solar metallicity. Based on our pPXF fitting over the stacked spectra of the extended LSB disk component (see Sec.~\ref{int-spat-ana} in Appendix), we obtain its mass-weighted mean stellar age as $\sim 7.4$ Gyr. Similarly, it is found as $\sim$ 5 Gyr old in the case of blueberry component. The age of recent starburst event observed with the H$\alpha$ based SFR of $\sim$ 17.4 M$_{\odot}~yr^{-1}$ is $\leqslant$ 10 Myr old. This age for recent starburst event is mainly constrained based on the detection of a broad Wolf-Rayet (WR) feature along with He{\sc ii} $\lambda$4686 emission only from the blueberry component \citep[see][]{Paswan2022}. The WR feature generally appears in a galaxy due to the presence a substantial ($10^2 - 10^5$) population of WR stars \citep[e.g.,][]{Kunth1981,Kunth1986} whose progenitors are evolved massive O-type stars. These O-type stars come to WR phase after $2 - 5$ Myr from their birth, spending a very short time in this phase \citep{Meynet2005}, before ending their life through supernova explosions. Therefore, the detection of WR phase constraints the age of recent starburst event in a galaxy. Overall, our stellar age analysis suggests that the blueberry galaxy under this study has experienced a multiple episodes of star formation, similar to typical class of dwarf galaxies \citep[e.g.,][]{Thuan1991,Krueger1995,Thornley2000,vanzee2001}. Furthermore, the detection of WR feature in our blueberry also suggests that this galaxy harbors a hard ionization radiation fields. This is consistent with our observed high [OIII]/[OII] emission line ratio (see Sect.~\ref{char}), showing one of the primary emission line properties of a typical CELLs galaxy \citep[e.g.,][]{Cardamoneetal2009,Yangetal2017}. \\

\begin{figure*}
\begin{center}
\rotatebox{0}{\includegraphics[width=0.33\textwidth]{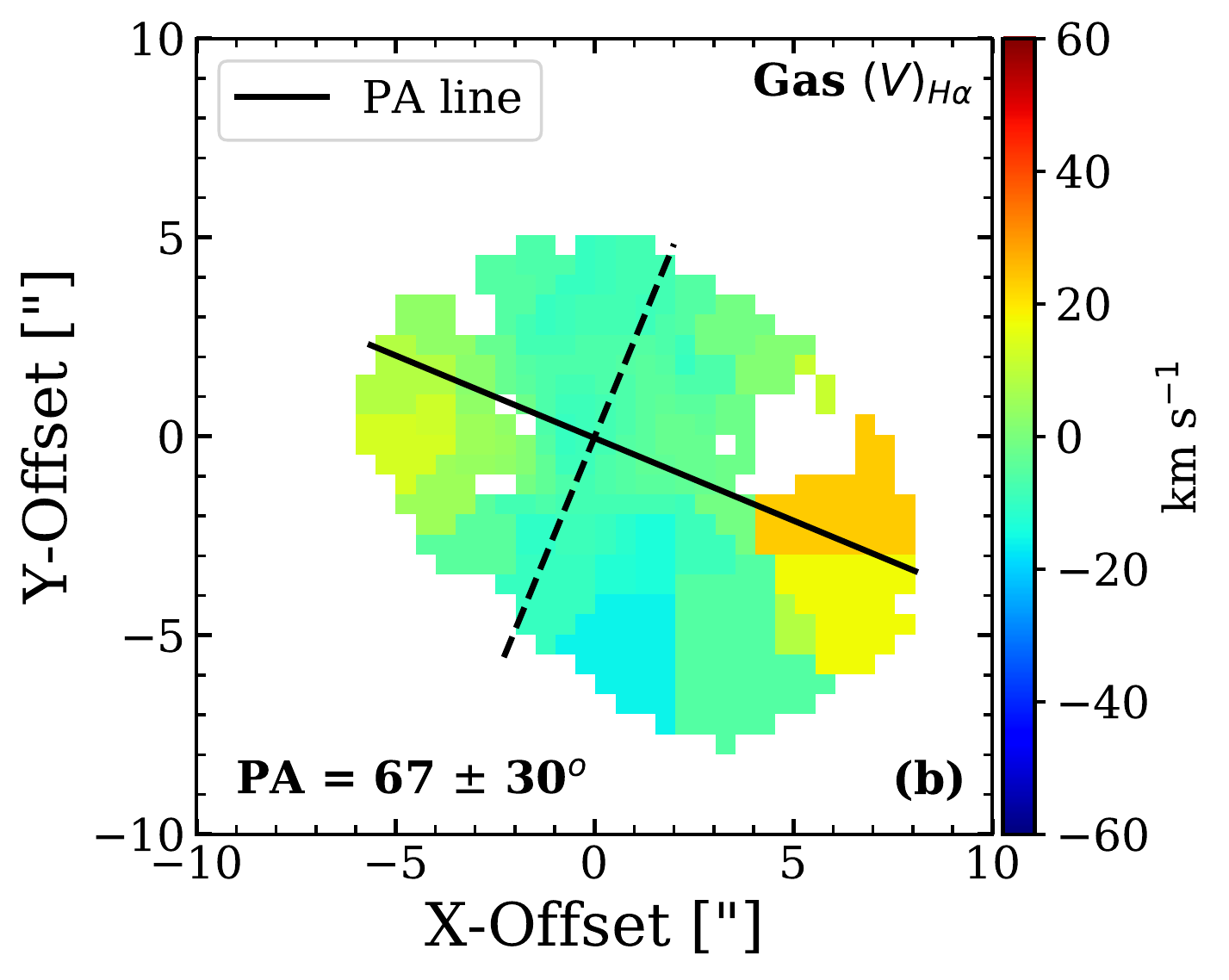}}
\rotatebox{0}{\includegraphics[width=0.32\textwidth]{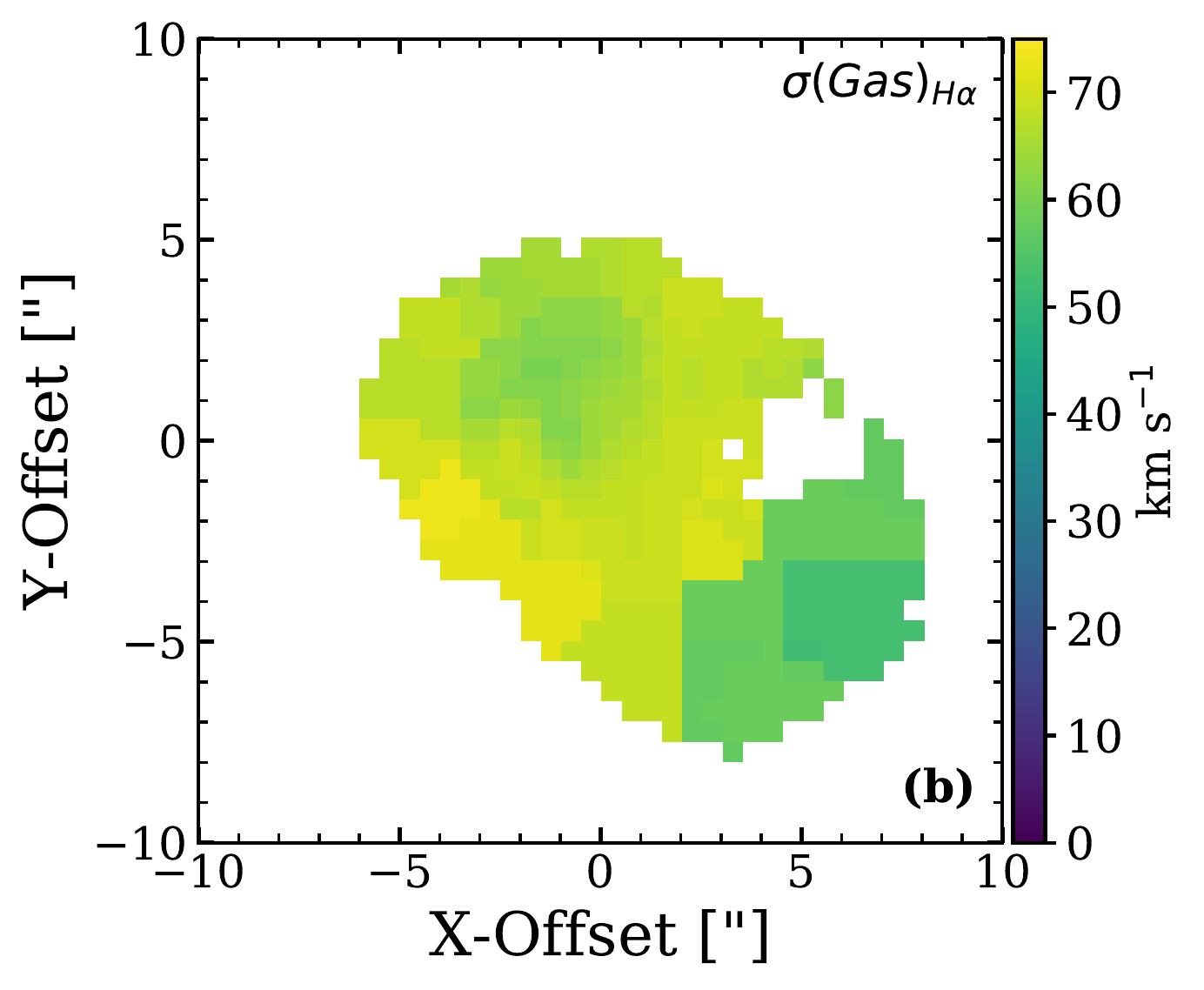}}
\rotatebox{0}{\includegraphics[width=0.31\textwidth]{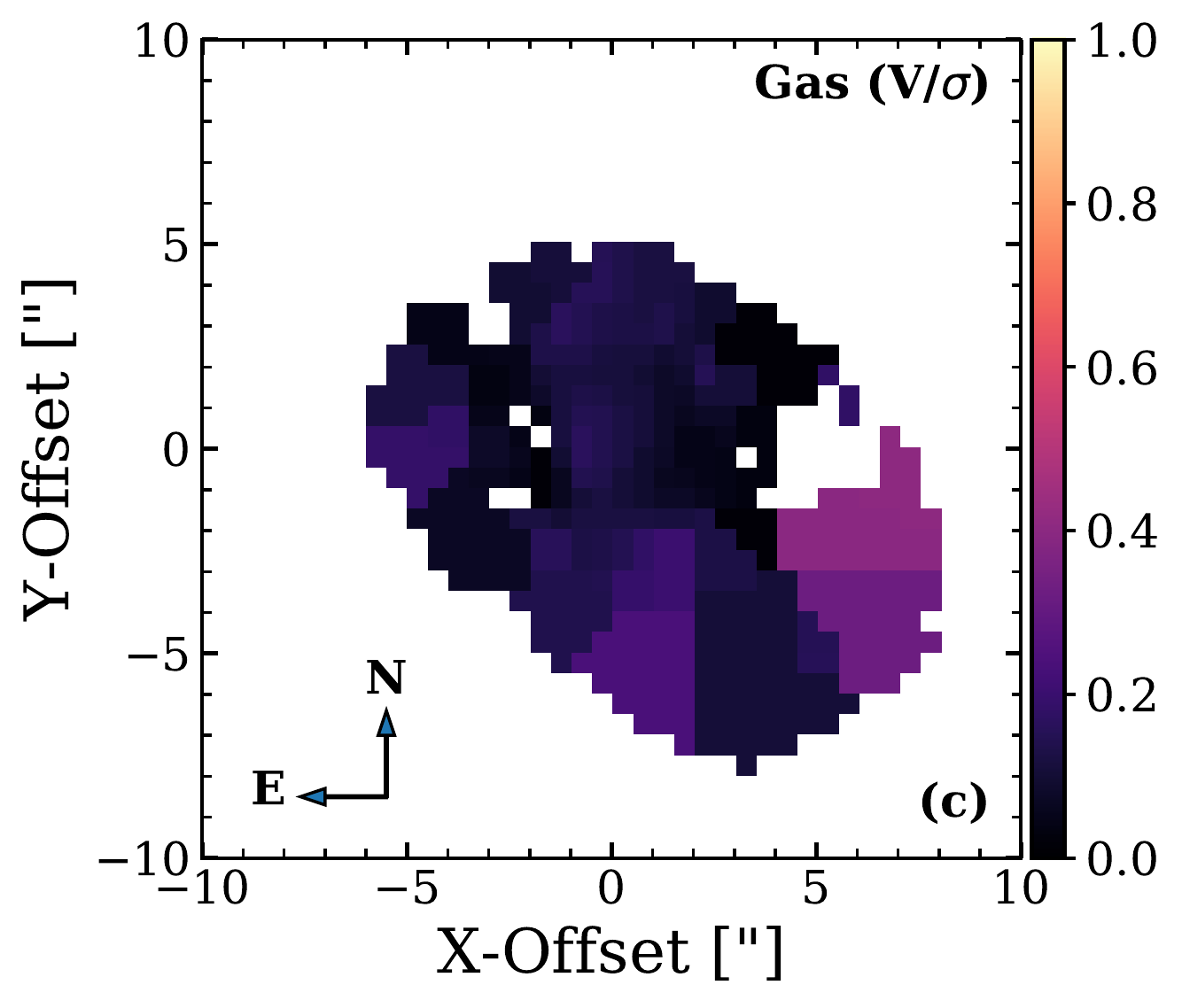}}
\caption{Plots representing the gas kinematics. The panel (a) shows rotation velocity of the ionized gas traced using H$\alpha$ emission line. In this panel, solid and dashed lines represent major and minor kinematic rotation axes of gas component, respectively. The panels (b) and (c) represent the gas velocity dispersion ($\sigma$) maps and $V/\sigma$ maps of gas, respectively.}
\label{fig:kinematics}
\end{center}
\end{figure*}

\section{Stellar and gas kinematics}
\label{kin}

The spatially-resolved kinematics of stars and gas in blueberry galaxies (or even in GPs) has largely remained unexplored, primarily due to the unavailability of IFU observations. In this line, MaNGA data of SHOC 579 represents the first IFU observation of a blueberry galaxy. The gas kinematics (see Fig.~\ref{fig:kinematics}) of the galaxy is derived using the strong H$\alpha$ emission line (corrected for instrumental broadening). However, only stellar kinematics could be obtained only in a limited sense using the stellar Mg{\sc i} $\lambda$5173,5167 absorption line. The full galaxy is divided into 11 bins each with $5\times 5$~spaxels. The left panel of Fig.~\ref{fig:MgI-stellar-kin} show the stacked spectra from each of these 11 bins. The bins marked by a3, a5, a9 a10 and a11 have $S/N \ge 3$. Note that the same was not possible with the stellar hydrogen absorption lines, because these lines were too weak to be detected in the spatially-resolved manner, even after stacking the spectra from 5 $\times$ 5 spaxels. Other stellar absorption features (e.g., Ca{\sc ii} $\lambda$8498,8542,8662 etc.) were not detected in the spectra.  \\

\par
We fit the Gaussian models to only those Mg{\sc i} $\lambda$5173,5167 absorption line which are detected with S/N $\textgreater$ 1.5 as shown in the left panel of Fig.~\ref{fig:MgI-stellar-kin}, and then derive the stellar line-of-sight velocity and dispersion. The derived stellar velocity dispersion is corrected for the instrumental broadening. The final stellar kinematics is shown in the right panel of Fig.~\ref{fig:MgI-stellar-kin}. Thanks to the MaNGA IFU observation and the proximity of SHOC 579 to us, it allowed us to derive the first stellar and gas kinematics of a blueberry galaxy in the literature. The left panels of Fig.~\ref{fig:MgI-stellar-kin} show the velocity, velocity dispersion and stellar $(v/\sigma)_{los}$ from each bin. From Fig.~\ref{fig:MgI-stellar-kin} and~\ref{fig:kinematics}, it can be seen that the identified LSB disk in our blueberry galaxy does not seem to have a well-defined rotation pattern, both in stars and gas. Although, the gas velocity pattern merely follows that of stars, gas seems to be rotating slower than the stars. They also appear to be misaligned by $\sim$ 30$^{o}$, estimated based on the PA fit module \citep{Cappellari2004}. The observed $(v/\sigma)_{los}$ map traced by the ionized gas suggest that the entire galaxy is rather pressure supported, similar to high-z galaxies \citep{Newmanetal2013} while the stellar $(v/\sigma)_{los} \simeq 1$ or so in the disk - a slight indication of a rotational support. It is worth mentioning here that prior to this study, \citet{Lofthouse2017} has reported the gas kinematics of four GPs, where they found two GPs with pressure supported kinematics and two with rotation supported.\\

In summary, our gross kinematic structure suggests a low angular momentum content in the galaxy; it might be possible that the disk has lost a good fraction of its angular momentum during the recent strong stellar feedback due to the starburst event \citep{FerraTolstoy2000, DOnghiaetal2006, Scannapiecoetal2008,Geneletal2015}. But this needs to be proven. On the other hand, it might be the case that we have been witnessing the recent kinematic-settling of an exponential LSB disk \citep{Kassinetal2012} in this low-mass starburst galaxy. Nevertheless, the readers may be cautioned that the finding of an exponential surface brightness profile (as shown in Sect.~\ref{sec:disk}) with mildly rotation supported stellar kinematics alone is not a compelling evidence of a thin stellar disk in dwarf systems. Because the possibility that the LSB host of SHOC579 is a triaxial stellar system, similar to other BCDs and dwarf ellipticals, cannot be discarded \citep[e.g.,][]{Sung1998,Papaderos2002}. In the literature, it is still a debated issue whether dwarf systems indeed host a stellar disk or triaxial stellar system. It is apparent that there is a need of high resolution IFU spectroscopy covering the full galaxy to establish the true kinematic state and whether the blueberry host contains a disk or a triaxial stellar system firmly.\\

 \section{Discussion and conclusions}
\label{sec:discuss}

In the present work, we show the structural properties of a special class of object so-called blueberry galaxy at $z \sim 0.0472$. Using its SDSS $i$-band image, we perform the GALFIT modeling that reveal an exponential LSB disk-like structure around blueberry region, presumably due to old stellar disk. However, we could not rule out if this is a generic property of nebular halo. With the help of MaNGA IFU data, our hypothesis for the presence of a faint old stellar disk is partially supported from the detection of Mg{\sc i} $\lambda$5173,5167  over the whole galaxy extent. The stellar kinematics indicates the presence of rotation support in the galaxy. Such kinematics with limited data quality, however, could not rule out if it is a triaxial stellar system. Overall, our results suggest that blueberry galaxy under this study has similar structural properties (see Fig.~\ref{fig:structure_compare}) and SFHs (containing stars from old, intermediate to young ages) as seen in other typical starburst dwarf galaxies e.g., BCDs \citep[e.g.,][]{Papaderos1996a,Papaderos1996b,Cairos2001,Noeske2003,GildePaz2005,Guseva2004,Papaderos2008,Thornley2000,vanzee2001}. Nevertheless, this class of objects is unique because of their extreme emission line properties and strong starburst events similar to high-redshift, compact, starburst LyC and Ly$\alpha$ emitters \citep[e.g.,][]{Jaskot2014,Henry2015,Yang2016,Huan2017,Verhamme2017,Izotov2021}.\\   
\par 

Our current results on blueberries, including work by \citet{Amorinetal2012} and \citet{Clarkeetal2021} on GPs, reveal that CELLs galaxies are luminous counterparts of local starburst dwarf galaxies such as BCDs. Therefore, the presence of LSB disk in these objects might be the general rule similar to BCDs, however, it is not clearly seen in most of the cases using ground-based SDSS-like survey. Perhaps, it became possible in our case due to proximity of our blueberry source (at $z \sim 0.047$) and its observed stellar mass of the LSB disk (i.e., $4.75\times 10^{9}$~M$_{\odot}$) whose surface stellar mass density is well above the SDSS detection limit. This detection limit is defined as $\sim$ 10$^{6}$ M$_{\odot}$~kpc$^{-2}$, assuming M$_{r}$/L$_{r}$ = 1  \citep{Kauffmannetal2003} with a typical integration time of $50 - 100$s \citep{Yorketal2000, Gunnetal2006}. \\

\begin{figure*}
\begin{center}
\rotatebox{0}{\includegraphics[width=0.7\textwidth]{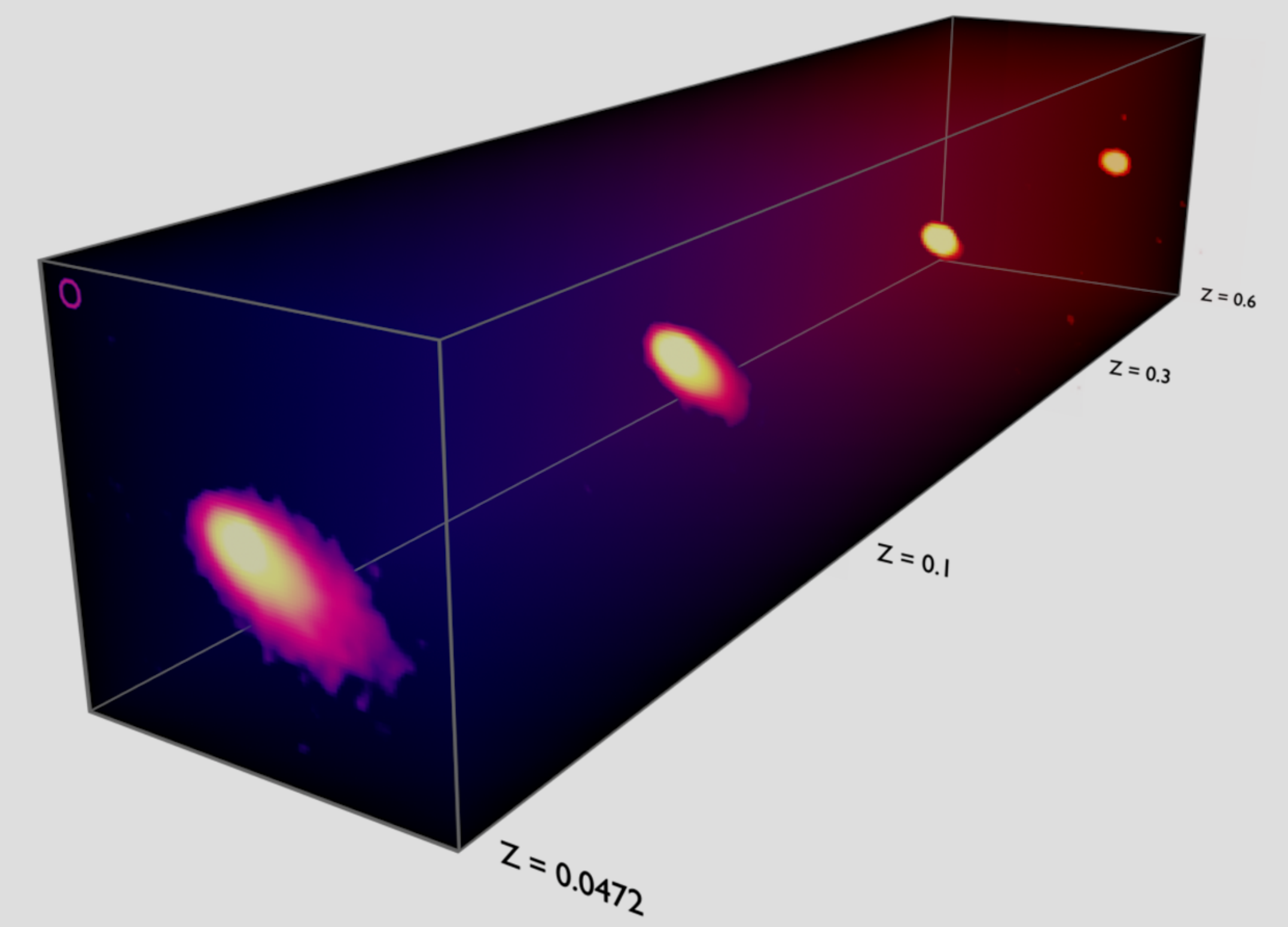}}
\caption{The 2D image visualization of our galaxy in this study at different redshifts from 0.0472 to 0.6. The circle shown at front upper left corner denotes the $SDSS$ (FWHM $\sim$ 1.2'') beam size.}
\label{fig:redshift-dim}
\end{center}
\end{figure*}

It is worth discussing here how a faint extended stellar disk in a CELLs galaxy situated at relatively higher redshift might be missed out in the SDSS-like surveys. For this, we performed an exercise on the dimming effect of our blueberry galaxy's image, after putting its 2D model images at various redshifts from z = 0.0472 to z = 0.6. In this exercise, at all the redshifts, we varied the apparent intensity of our source by a factor depending upon redshift as (1+z)$^{-4}$, an effect known as redshift dimming or Tolman dimming \citep[see][]{Tolman1930,Tolman1934,Tolman1935}. In the redshifted model images, we add Poisson noise in combination with the SDSS background noises so that the model images mimic the observed SDSS images. Note that as the source moves towards higher redshifts its angular size will vary following the cosmological scale parameter in the units of kpc/arcsec. Assuming the SDSS plate scale \citep[i.e., 0.396 arcsec/pixel;][]{Gunn1998} and physical size of the source in the units of kpc, we have taken care of the variation in angular size of the galaxy images at different redshifts. A visual demonstration of the redshift dimming of our galaxy is presented in Fig.~\ref{fig:redshift-dim}. It can be seen that the underlying LSB disk starts disappearing into the background noise beyond redshift 0.1 or so. This implies that if we place our blueberry galaxy at $z > 0.1$, we may not be able to resolve its two components (i.e., a disk and inner Sersic component) using the SDSS-like data, having a beam size of 1.2'' FWHM \citep[see][]{Ross2011}. Our experiment indicates that if a faint LSB old stellar disk in GPs at relatively higher redshifts is indeed very common, it can be detected only with very sensitive and high angular resolution observations performed using either bigger ground or space-based telescopes. \\

\par
The extreme emission line nature of CELLs galaxies originates during their recent strong starburst event. The mechanism responsible for triggering such strong starburst events in these systems is remains unexplored. In this context, stellar and gas kinematics of our blueberry galaxy can shed some light on this aspect. Our blueberry galaxy has disturbed kinematics, showing a misalignment of $\sim$ 30$^{o}$ between the stellar and gaseous kinematic major axes. In general, such kinematic misalignment in galaxies indicates signs of galaxy interaction and merger or external gas accretion \citep{Jaiswal2013,Voort2015,Jin2016,Paswan2018,Song-lin2021}.
Furthermore, the study by \citet{Lofthouse2017} has also shown an evidence of minor-merger in their two GPs. Recently, \citet{Kanekar2021} presented the estimates of H{\sc i} gas mass in 40 GPs observed with single radio dish Arecibo Telescope and Green Bank Telescope (GBT). Interestingly, they found that nine of their 40 GPs lie above ($\textgreater$ 0.6 dex) the local M$_{HI}$ - M$_{B}$ relation \citep[e.g.,][]{Denes2014}, suggesting that these nine GPs are gas-rich for their given optical luminosity and might be experiencing external gas accretion or minor merger. From our above discussion, it seems that an external gas accretion via galaxy interaction or minor merger might be one of the potential causes for triggering starburst events in the CELLs galaxies. Nevertheless, this scenario needs to explore further in grater depth using a large sample of CELLs galaxies. \\
\par 
It is important to point out here that - a good fraction of CELLs galaxies have already been identified as LyC leakers and/or Ly$\alpha$ emitters \citep[e.g.,][]{Izotovetal2016,Izotovetal2018,Izotovetal2018b,Huan2017,Verhamme2017,Izotov2021}, mimicking their high redshift counterparts that have contributed to the re-ionization process at $z \geq 6$. However, the host properties of these local counterparts show the presence old stellar population extended to a disk-like structure, unlike their high-redshift counterparts. The host properties of high-z galaxies are expected to be very young (maybe having their first generation stars) and compact without old stellar extended disk-like structure. It means that mechanisms that support the escape of ionizing photons in the local CELLs galaxies may be different from those at play during the epoch of re-ionization. Along this line, the upcoming JWST would play a vital role in establishing the true underlying nature of the stellar population in the high-redshift galaxies; at the same time establishing a possible connection with blueberries like SHOC 579 studied here.\\
 
Our main conclusions are:

\begin{itemize}
    \item {The blueberry source studied here has recently ($\leqslant$ 10 Myr) formed over an underlying LSB disk of its host galaxy. To date, this blueberry galaxy is the most massive and metal-rich one around which we have the direct observational evidence of an old stellar population.
    
    \item{Based on the modelling of stacked spectra showing clear detection of Mg{\sc i} $\lambda$5173,5167 absorption line, we find the average mass-weighted age of the stellar population to be $\sim 5$~Gyr and $\sim 7$~Gyr for the blueberry component and the stellar disk, respectively. These estimates are consistent with our results from the best-fit SED model.}
    
    \item This work reveals the first kinematics of stars and gas of a blueberry galaxy based on locally stacked spectra from MaNGA IFU observation. In that, the studied blueberry galaxy is found to be dispersion dominated in the ionized gas component and mildly rotation supported in stellar component. Furthermore, these stellar and gaseous components are misaligned to each other.}
    
    \item {The structural and SFH analyses of the host galaxy of our blueberry in this work and GPs presented by \citet{Amorinetal2012} and \citet{Clarkeetal2021} in the literature suggest that the CELLs galaxies (i.e., blueberries and GPs) do not represent peculiar cases of dwarf galaxy evolution. In fact, with respect to the structural properties of their host galaxies, they are compatible with a common evolutionary track of the main population of BCDs. In other words, the CELLs galaxies are luminous counterparts of local low-luminous BCDs. Their extreme emission line properties are likely due to recent strong starburst events, potentially triggered by external gas accretion process. 
    
    \item As such, CELLs galaxies are often referred to as the best analogs of the sources at epoch of re-ionization ($z \textgreater 6$). However, the above conclusion revealing the presence old stars in the CELLs galaxies would imply that $-$ mechanisms that allow escape of ionizing photons in these local objects may be different from those at play during the epoch of re-ionization. Exploring the physical process that drives the escape of ionizing photons in both local and high redshift LyC leakers is very important.
    }
    
\end{itemize}

\section*{Acknowledgments}

We thank the anonymous referee for his/her constructive suggestions that have greatly improved the contents and quality of the paper. AP thanks R. Amorin and P. Papaderos for kindly providing the data points used in Fig.~\ref{fig:structure_compare}. AP also thanks Kuldeep Singh for his help in making of Fig.~\ref{fig:redshift-dim}. This research has made use of the SAO/NASA Astrophysics Data System (ADS) operated by the Smithsonian Astrophysical Observatory (SAO) under a NASA grant.\\

\software{SExtractor \citep{Bertin1996}, GALFIT \citep{Pengetal2002}, IRAF \citep{Tody1986}, PHOTUTIL \citep{Larry2020}, ASTROPY \citep{price2018astropy}, PCIGALE \citep{Boquienetal2019}, DRP \citep{Lawetal2016}, DAP \citep{Westfalletal2019}, pPXF \citep{Cappellari2004}}

\appendix

\section{PSF matched photometry}
\label{psf-match}

Total magnitude of the central blueberry region and the full galaxy are estimated using aperture photometry technique. Since the images from FUV to IR bands are taken from different survey/instruments whose PSFs are different to each other, we therefore applied a PSF matching technique, before estimating the fluxes of blueberry region and full galaxy. In this technique, we first estimated various PSF kernals between the $GALEX$ FUV image (having the largest PSF among all the bands) and the rest of other band images (i.e., from optical to IR bands that have a better PSF compared to FUV band) using PHOTUTIL \citep{Larry2020} task available in ASTROPY \citep{price2018astropy} package. We did not apply the PSF matching on the $GALEX$ NUV band image as its PSF is very close to the FUV band image. After obtaining various PSF kernals, we convolved them over their respective band images from optical to IR. This process provides us with all the band images from FUV to IR having a similar PSF. We then measured the aperture size of the central blueberry region only and the full galaxy in the SDSS $i$-band, which are respectively found to be $\sim$ 2 and 8 arcsec in radius. In rest of the band images, we used the same aperture sizes for the central blueberry region only and the full galaxy. Finally, we estimated the magnitudes of the galaxy in these two apertures separately using all the PSF matched band images from FUV to IR. Note that in each case the magnitude of central blueberry region is estimated after subtracting background contamination from the surrounding LSB envelope, where the background contamination is measured using an annular region of 3 pixels around central blueberry region.

\section{SED modeling}
\label{sed}

We constructed multi-band SED for the full galaxy and the central blueberry region separately using archival data from FUV to IR bands. The fluxes for SED construction are obtained from our PSF-matched photometry as explained in the above section. Some of the physical properties of full galaxy and blueberry component have been derived by fitting the stellar population model with nebular lines using PCIGALE \citep{Boquienetal2019}. In that, we use BC03 stellar population library \citep{Bruzual2003} with exponentially declining star formation histories with late bursts having a Salpeter Initial Mass Function \cite[IMF;][]{Salpeter1955} with lower and upper mass cutoffs at 0.1 and 100 M$_{\odot}$ respectively. While performing the SED modelling, we have used exponentially declining star-formation history (SFH) with late burst. The e-folding timescale of the main stellar population is varied from $\tau_{main} = 200, 250, 300, 350, 400, 500, 600, 700, 1000, 1500, 2000$ Myr after some experimentation. The e-folding timescale of the late burst, $\tau_{burst}=20., 30, 40, 50, 70, 100, 200$~Myr. Similarly, the age of the main stellar population has also been varied from $1000 - 9000$~Myr, in steps of 200 Myr upto 2000 Myr and 500 Myr beyond 2000 Myr. Initially, we have varied metallicity values from $0.004, 0.008, 0.02$.
The metallicity values for the best-fit models for the full galaxy and the blueberry region are found to be $0.02$ and $0.008$ respectively. These values are in accordance with our measurements from the MaNGA IFU data. 
The values of color excess, $E(B - V)$, are used in the range of $0.1 - 0.2$, guided by our measurements using the Balmer decrement from MaNGA spectra. For the reduction factor for the old stellar population, we followed the Calzetti relation $E(B - V)_{star} = 0.44 \times E(B - V)_{nebular}$. We apply Calzetti extinction law \citep{Calzetti2000} for dust modeling. The dust attenuation curve has a UV bump at 2175~\AA~ with amplitude $\sim 1/3$ of that of the MW bump and the overall power-law slope of the curve is fixed at $n=-\delta +0.75$, where $\delta$ is the slope deviation. In our modelling, we have varied $\delta$ from -0.5 to -0.1 in steps of 0.1. For the best-fit model ($\chi^2_{reduced,full}=8.0$ and $\chi^2_{reduced,BBonly}=5.2$) presented in Fig.~\ref{fig:MS}, $\delta=-0.25$ for the Blueberry and $-0.2$ for the full galaxy model. These slopes are similar to the slope of the SMC extinction curve (although slightly shallower) which may be more appropriate for high redshift SFGs \citep{Salim2018} and their local analogs.
We obtain $\tau_{main}=250$~Myr, total stellar mass $M_{*}=5.2\times10^9$~M$_{\odot}$ and mass-weighted stellar age of $7.05$~Gyr. While for the Blueberry component, we obtain stellar mass as $M_{*}=4.5\times10^8$~M$_{\odot}$ with a mass-weighted age of the stellar population to be $5.6$~Gyr. The central blueberry component had $\tau_{burst}=100$~Myr and age of the starburst population $burst_{age}=10$~Myr. In addition to $\chi^2$ as a goodness of fit, we have also paid attention to comparing and picking the best SED models which are able to explain the observation better. For example, we have compared the ratio of rest-frame equivalent width EW of strong emission lines such as EW({O[{\sc iii}]})/EW({H{$\alpha$}}). This ratio is 0.96 from the observed MaNGA spectra while this ratio is 1.1 from our best-fit SED model of the blueberry component. The mass-weighted stellar population age from our best-fit model matches closely with that obtained from the pPXF modelling of the stacked spectra (see Appendix~\ref{int-spat-ana} below). 

\section{Galfit modeling} \label{Galfit}

We discuss here the 2D modeling of our studied galaxy's light distribution in the SDSS $i$-band using GALFIT tool provided by \citep{Pengetal2002}. For the GALFIT modeling of our galaxy, we compile the Gaussian PSF model obtained after creating a PSF kernal model image with 1.06" FWHM. This PSF size for the SDSS $i$-band image is taken as per the information provided in the SDSS Science Archive Server. With this, we first proceed with a simple two component fitting - an inner Sersic component surrounded by an exponentially declining disk. The first GALFIT run on the full galaxy using two components and without any constraints resulted in a significantly off-centered disk towards the South-West of the galaxy. Such a highly off-centered disk is unphysical. In subsequent GALFIT run, we therefore constrained both the components to have a common center. This approach yet yielded an improper model which is confirmed by an inspection of its residual image (i.e., observed minus model). To obtain the best fit model to the full galaxy, we applied a different approach $-$ we first model the underlying disk component, and then the inner Sersic component. Prior to fitting the exponential disk, we masked the bright inner component of the galaxy with a circular mask of $\sim$2 arcsec in radius as shown in Fig. \ref{galfit}(b). This masking ensures that any effect due to central bright region of the galaxy is minimised, and also does not rule out the possible existence of an off-centered disk, if it indeed exists.\\

Since our preliminary visual inspection of the galaxy shows the presence of a sheared disk-like light distribution around the central blueberry component - indicating a possible lopsided disk, we therefore modified the pure exponential disk function as given by Eq.~\ref{eq-1}1 into Eq.~\ref{eq-2}2, after including the first order Fourier mode as follows: 

\begin{equation}
  I(x,y) = I_{0}~exp\Bigg(\frac{-r(x,y)}{r_{h}} \Bigg)
 \end{equation}\label{eq-1}
 
 \begin{equation}
  I(x,y) = I_{0}~exp\Bigg(
 \frac{-r(x,y) \times~\Big(1 + \sum^{N}_{m=1} a_{m}~cos~\Big( m(~\theta + \phi_{m}) \Big) }{r_{h}}~\Bigg)
  \end{equation}\label{eq-2}

In the above equations, $I_{0}$ and $r_{h}$ are peak intensity and scale-length of the exponential disk respectively, and $m = 1$ represents the first order Fourier mode. $r(x,y)$ represents the radial coordinates of a standard ellipse which is defined as :

 \begin{equation}
  r(x,y) = \Bigg( |x-x_{0}|^{2} + \Big| \frac{y-y_{0}}{q} \Big|^{2}~\Bigg)^{\frac{1}{2}}
  \end{equation}

Here, $q$ and $a_{m}$ are axis ratio and Fourier amplitude respectively. ($\theta~+~\phi_{m}$) represents the relative angle between mode $m$ and position angle (PA) of the standard ellipse \citep{Pengetal2010}, where  the $\theta$ is defined as tan$^{-1}((y-y_{0})~/~(x-x_{0})q)$. Using the model as described above, we found a better fit to the underlying disk as confirmed by obtaining the best residual image. We then fit a simple Sersic component to the obtained residual image (see Fig.~\ref{galfit}d). The complete model of our galaxy reveals that the centers of two components are indeed not common $-$ indicating the presence of an off-centered disk. The final GALFIT model of the full galaxy and its residual are shown in Fig.~\ref{galfit} (c) and (f), respectively. We further analysed the statistics of residual image using the distribution of pixel values within the dashed-box as shown in Fig.~\ref{galfit}(f). This distribution follows a normal distribution function whose $\mu$ and $\sigma$ are found to be $\sim$ 0.0034 and $\sim$ 0.0542 nmgy, respectively. Finally, the GALFIT modeled images and fitted parameters are presented in Fig.~\ref{galfit} and Table~\ref{tab:gal-params}, respectively. \\

\begin{table*}[!h]
      \centering
      \caption{The best output parameters obtained from the GALFIT fitting to the full galaxy.}
      \begin{tabular}{ccc}
      \hline\hline
          Parameters & Sersic component & Exponential component  \\
           \hline
         center (x,y)  & (50.27$\pm$0.06,  50.75$\pm$0.07) & (46.63$\pm$0.63,  51.68$\pm$0.49) \\
         Integrated Magnitude (AB)  & 17.63$\pm$0.03 & 17.65$\pm$0.05 \\
         Scale-length (kpc) & --& 1.54$\pm$0.13
         \\
         Effective radius (pc)  & 217$\pm$51 & --\\
         Sersic Index  & 1.48 $\pm$1.21 & --\\
         Axis ratio (b/a)  & 0.69$\pm$0.21 &0.66$\pm$0.03 \\
         Position Angle (deg) & 64$\pm$20 & 69$\pm$5 \\
         Fourier Amplitude  & -- & -0.57$\pm$0.05 \\
         Fourier mode phase angle (deg) & -- &  -11$\pm$12 \\
         \hline
      \end{tabular}
      \label{tab:gal-params}
\end{table*}

\section{Integrated spectral analyses}\label{int-spat-ana}

\begin{figure*}[!h]
\begin{center}
\rotatebox{0}{\includegraphics[width=0.47\textwidth]{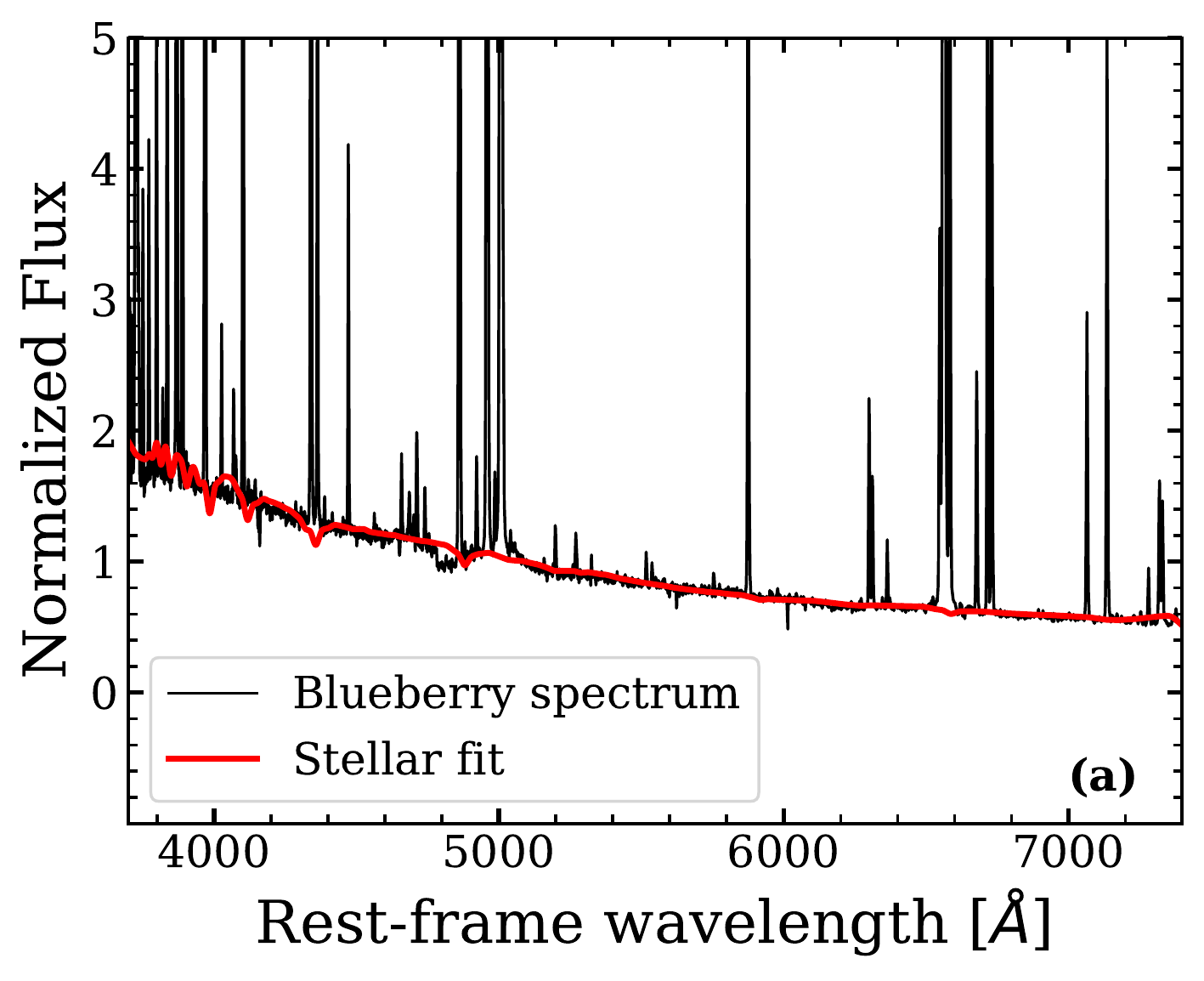}}
\rotatebox{0}{\includegraphics[width=0.47\textwidth]{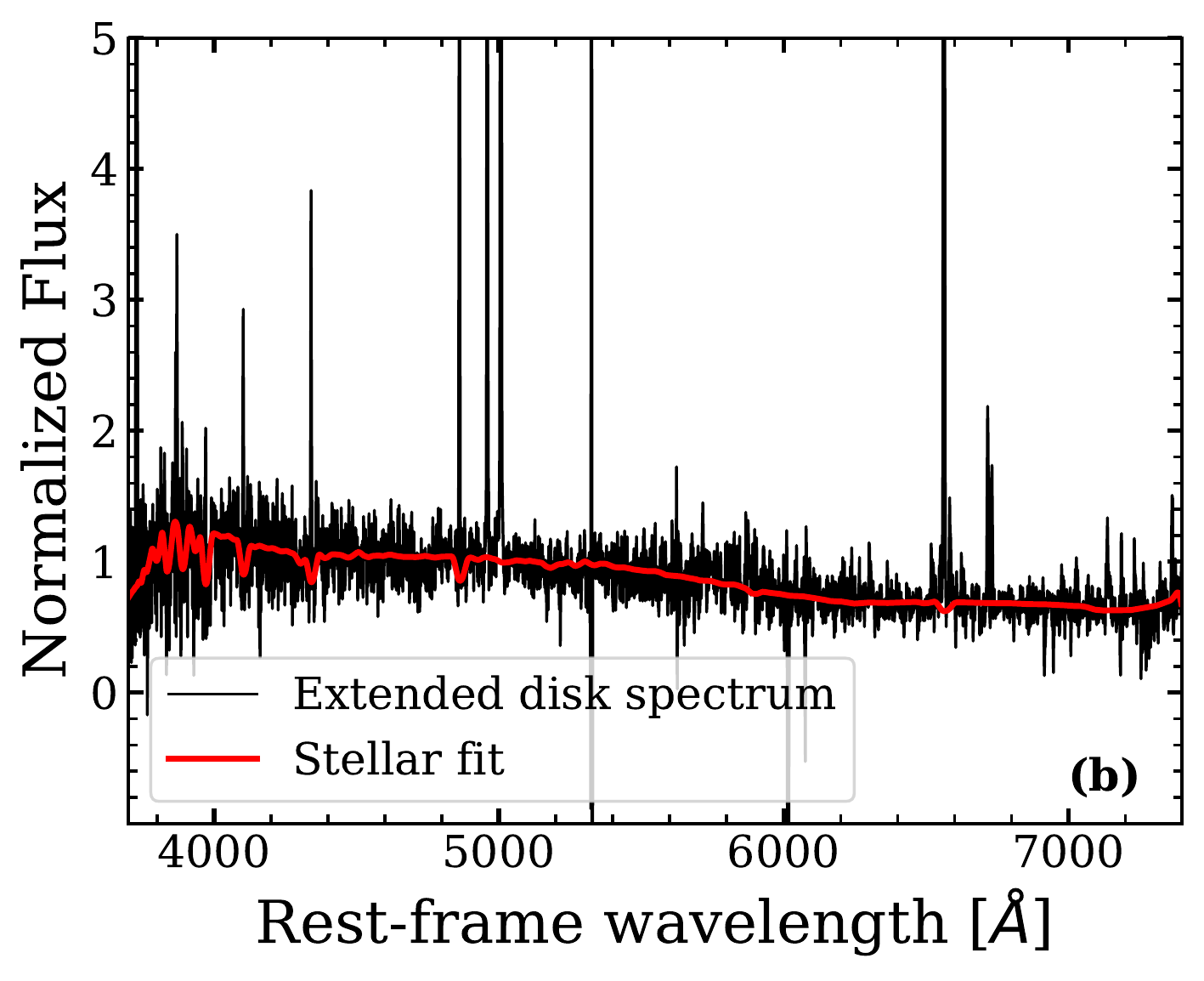}}
\caption{The stellar-model fit (red line) to the observed stacked spectra of (a) central blueberry region and (b) outer LSB stellar disk. These models are fitted using pPXF code with MILES stellar libraries. In each case, the observed spectra are shown by black lines.}
\label{fig:ppxf-fit}
\end{center}
\end{figure*}

In our analysis, we noticed that the spectral features related to underlying old stellar population such as Balmer  and MgI $\lambda$5173 absorption lines are too faint to detect in the spatially resolved manner using the MaNGA data cube. We therefore performed the spectra stacking using a few hundreds of spectra from both the central blueberry component and surrounding faint LSB disk regions. The stacked spectra from these two regions are shown in Fig.~\ref{fig:ppxf-fit} (a) and (b), where red stellar continuum fittings are drawn using pPXF code with MILES stellar libraries \citep{Cappellari2004}. These fittings clearly indicate the presence of underlying stellar absorption features, and lead to the estimates of mean stellar ages as $\sim$ 5 Gyr and $\sim$ 7.4 Gyr, respectively, for blueberry component and faint LSB disk regions.

\bibliography{ref}

\end{document}